\documentclass[oneside,american,canadian,oldfontcommands,10pt,onecolumn,oneside,a5paper,article,extrafontsizes,sort&compress,numbers]{memoir}
\usepackage[LGR,T1]{fontenc}
\pdfoutput=1
\usepackage[utf8]{inputenc}
\usepackage[a5paper]{geometry}
\usepackage{babel}
\usepackage{prettyref}
\usepackage{float}
\usepackage{textcomp}
\usepackage{mathtools}
\usepackage{bm}
\usepackage{amsmath}
\usepackage{amsthm}
\usepackage{amssymb}
\usepackage{graphicx}
\usepackage{esint}
\usepackage[numbers]{natbib}
\OnehalfSpacing
\usepackage[unicode=true,
 bookmarks=true,bookmarksnumbered=false,bookmarksopen=false,
 breaklinks=true,pdfborder={0 0 1},backref=false,colorlinks=false]
 {hyperref}
\hypersetup{pdftitle={Pairwise maximum-entropy models and their Glauber dynamics: bimodality, bistability, non-ergodicity problems, and their elimination via inhibition},
 pdfauthor={Vahid Rostami, PierGianLuca Porta Mana , Moritz Helias},
 unicode=true,bookmarks=true,bookmarksnumbered=false,bookmarksopen=false, breaklinks=true,pdfborder={0 0 0},backref=false,colorlinks=false,dvipdfm}

\makeatletter

\DeclareFontEncoding{LGR}{}{}
\DeclareTextSymbol{\~}{LGR}{126}

\usepackage{enumitem}		% customizable list environments

\newcommand{\pdfauthor}{V. Rostami, P.G.L. Porta Mana, M. Helias}
\newcommand{\headtitle}{Bimodality \& inhibition in maximum-entropy}% title for headers
\newcommand{\headauthor}{Rostami, Porta Mana, Helias}% author for headers

\copypagestyle{customheader}{plain}
\makeheadrule{customheader}{\headwidth}{0.5\normalrulethickness}
\makeoddhead{customheader}%
{\footnotesize\sffamily\scshape\headauthor}{}{\footnotesize\sffamily\headtitle}
\makeoddfoot{customheader}{}{\thepage}{}
\hypersetup{colorlinks=false,%
bookmarksnumbered,%
pdfborder={0 0 0.25},%
citebordercolor={0.2 0.1333 0.5333},%bluish
linkbordercolor={0.0667 0.4667 0.2},%greenish
urlbordercolor={0.5333 0.1333 0.3333},%reddish
breaklinks=true,%
pdfauthor = {\pdfauthor}%
}

\captiontitlefont{\footnotesize}
\captionnamefont{\footnotesize\bfseries}

\usepackage{ifthen}
\newboolean{isarxiv}

\usepackage{lmodern}
\usepackage[utf8]{inputenx}

\usepackage{textcomp}

\usepackage{upgreek}

\usepackage{bm}
\usepackage{amsmath,amssymb}

\usepackage{nameref}

\usepackage{microtype}
\usepackage{rotating}

\makeatletter
\renewcommand{\@biblabel}[1]{\quad#1.}
\makeatother

\date{16 May 2016}

\usepackage{graphicx}

\usepackage{prettyref}

\newrefformat{cap}{{fig.~\ref{#1}}}
\newrefformat{fig}{{fig.~\ref{#1}}}
\newrefformat{tab}{{table ~\ref{#1}}}
\newrefformat{sec}{{§~\ref{#1}}}
\newrefformat{sub}{{§~\ref{#1}}}
\newrefformat{eq}{{eq.~\eqref{#1}}}

\chapterstyle{reparticle}
\counterwithout{section}{chapter}

\usepackage{etoolbox}
\makeatletter
\patchcmd{\@startsection}{\@ssect{#3}{#4}{#5}{#6}}{\@dblarg{\@sect{#1}{\@m}{#3}{#4}{#5}{#6}}}{}{\PackageError{fix-unnumbered-sections}{Unable to patch \string\@startsection; are you using a non-standard document class?}\@ehd}
\makeatother

\pretitle{\begin{center}\Large\sffamily\bfseries}
\posttitle{\bigskip%
\end{center}}

\preauthor{\vspace{-0\jot}\begin{center}
\normalsize\sffamily\lineskip  0.5em%
\begin{tabular}{c}}
\postauthor{\end{tabular}\par\end{center}}

\predate{\bigskip\begin{center}\small}
\postdate{\end{center}}

\newcommand*{\customabstract}{%
\abslabeldelim{}%
\renewcommand*{\absnamepos}{flushleft}
\setlength{\absleftindent}{0pt}%
\setlength{\absrightindent}{0pt}%
}

\setfloatadjustment{figure}{\footnotesize\centering}
\firmlists*
\midsloppy
\DeclarePairedDelimiter\expp{(}{)}

\DeclarePairedDelimiter\set{\{}{\}}

\providecommand{\foreignlanguage}[2]{#2}

\makeatother

\begin{document}
\global\long\def\second{\,\mathrm{s}}
\global\long\def\ms{\,\mathrm{ms}}
\global\long\def\minute{\,\mathrm{min}}

\global\long\def\e{\mathrm{e}}
\global\long\def\di{\mathrm{d}}
\global\long\def\de{\mathrm{\partial}}

\global\long\def\yss{\bm{s}}
\global\long\def\yst{S}
\global\long\def\ysm{\bar{s}}
\global\long\def\ysmin{\bar{s}_{\text{m}}}
\global\long\def\ysi{s_{\text{I}}}
\global\long\def\ysms{\bar{s}_{\text{sub}}}

\global\long\def\ysa{s'}
\global\long\def\yra{q'}

\global\long\def\hh{\bm{\mu}}
\global\long\def\yhhm{\bar{\mu}}
\global\long\def\yhha{\mu'}
\global\long\def\yhhr{\mu_{\text{r}}}

\global\long\def\jj{\bm{\varLambda}}
\global\long\def\yjjm{\bar{\varLambda}}
\global\long\def\yjja{\varLambda'}
\global\long\def\yjjr{\varLambda_{\text{r}}}
\global\long\def\yji{\varLambda_{\text{I}}}
\global\long\def\yji{\varLambda_{\text{I}}}

\global\long\def\PP{P_{\text{p}}}
\global\long\def\PR{P_{\text{r}}}
\global\long\def\PI{P_{\text{i}}}
\global\long\def\zp{Z_{\text{p}}}
\global\long\def\zr{Z_{\text{r}}}
\global\long\def\zi{Z_{\text{i}}}

\global\long\def\PT{P_{\text{t}}}

\global\long\def\mm{\bm{m}}
\global\long\def\ymmm{\bar{m}}

\global\long\def\yccm{\bar{c}}
\global\long\def\cc{\bm{c}}
\global\long\def\yrhom{\bar{\rho}}
\global\long\def\rrho{\bm{\rho}}

\global\long\def\ygg{\bm{g}}
\global\long\def\yggm{\bar{g}}
\global\long\def\yG{G}
\global\long\def\HS{\mathrm{H}}
\global\long\def\dirac{\updelta}

\global\long\def\expe#1{\mathrm{\mathrm{E}}\expp{#1}}
\global\long\def\expeb#1{\mathrm{\mathrm{E}}\expp*{#1}}

\global\long\def\expep#1{\mathrm{\mathrm{E}}_{\text{p}}\expp{#1}}
\global\long\def\expebp#1{\mathrm{\mathrm{E}}_{\text{p}}\expp*{#1}}

\global\long\def\exper#1{\mathrm{\mathrm{E}}_{\text{r}}\expp{#1}}
\global\long\def\expebr#1{\mathrm{\mathrm{E}}_{\text{r}}\expp*{#1}}

\global\long\def\expei#1{\mathrm{\mathrm{E}}_{\text{i}}\expp{#1}}
\global\long\def\expebi#1{\mathrm{\mathrm{E}}_{\text{i}}\expp*{#1}}

\global\long\def\expet#1{\mathrm{\mathrm{E}}_{\text{t}}\expp{#1}}
\global\long\def\expebt#1{\mathrm{\mathrm{E}}_{\text{t}}\expp*{#1}}

\global\long\def\tav#1{\widehat{#1}}
\global\long\def\av#1{\overline{#1}}

\global\long\def\T{^{\intercal}}
\global\long\def\cond{\mathpunct|}
\global\long\def\dotv{\mathord\cdot}
\global\long\def\Land{\mathbin{\ \land\ }}
\global\long\def\ynu{\nu}

\title{Pairwise maximum-entropy models\\ and their Glauber dynamics:\\ bimodality, bistability, non-ergodicity problems,\\ and their elimination via inhibition}

\author{Vahid Rostami \textsuperscript{1{*}}, PierGianLuca Porta Mana \textsuperscript{1},
Moritz Helias \textsuperscript{1,2}}

\maketitle
\noindent {\scriptsize{}\textsuperscript{1} Institute of Neuroscience
and Medicine (INM-6) and Institute for Advanced Simulation (IAS-6)
and JARA BRAIN Institute I, Jülich Research Centre, Germany}\\
{\scriptsize{}\textsuperscript{2} }\foreignlanguage{american}{{\scriptsize{}Department
of Physics, Faculty 1, RWTH Aachen University, Germany}}{\scriptsize \par}

\noindent {\scriptsize{}\textsuperscript{{*}} \href{mailto:v.rostami@fz-juelich.de}{v.rostami@fz-juelich.de}}{\scriptsize \par}

\bigskip{}

\iffalse%\pdfbookmark[1]{Title}{TitlePage}

\let\oldnameref\nameref \renewcommand{\nameref}[1]{\textit{``\oldnameref{#1}''}}

%\setboolean{isarxiv}{true}

%\vspace*{0.35in}

%\begin{flushleft}

% Title must be 250 characters or less. % Please capitalize all terms in the title except conjunctions, prepositions, and articles. %\begin{flushleft}
{
\Large \textbf
\newline
{Pairwise maximum-entropy models and their Glauber dynamics:\\
bimodality, bistability, non-ergodicity problems, and their elimination via inhibition} 
\addcontentsline{toc}{section}{Title}
}\newline
\\
Vahid Rostami \textsuperscript{1*},
PierGianLuca Porta Mana \textsuperscript{1},
%Sonja Gr\"un \textsuperscript{1,3}, and
Moritz Helias \textsuperscript{1,2}
\\
% with the Lorem Ipsum Consortium\textsuperscript{\textpilcrow} \\
\bigskip
{\footnotesize
{\bf 1} Institute of Neuroscience and Medicine (INM-6) and Institute for Advanced Simulation (IAS-6) and JARA BRAIN Institute I, J\"ulich Research Centre,  Germany\\
{\bf 2} Department of Physics, Faculty 1, RWTH Aachen University, Germany

\bigskip

* \href{mailto:v.rostami@fz-juelich.de}{v.rostami@fz-juelich.de}

\end{flushleft}
\fi
\thispagestyle{plain}
%\maketitle
\customabstract

\begin{abstract}
{\footnotesize{}}{\footnotesize \par}

{\footnotesize{}Pairwise maximum-entropy models have been used in
recent neuroscientific literature to predict the activity of neuronal
populations, given only the time-averaged correlations of the neuron
activities. This paper provides evidence that the pairwise model,
applied to experimental recordings, predicts a bimodal distribution
for the population-averaged activity, and for some population sizes
the second mode peaks at high activities, with $90\%$ of the neuron
population active within time-windows of few milliseconds. This bimodality
has several undesirable consequences: 1.~The presence of two modes
is unrealistic in view of observed neuronal activity. 2.~The prediction
of a high-activity mode is unrealistic on neurobiological grounds.
3.~Boltzmann learning becomes non-ergodic, hence the pairwise model
found by this method is not the maximum entropy distribution; similarly,
solving the inverse problem by common variants of mean-field approximations
has the same problem. 4.~The Glauber dynamics associated with the
model is either unrealistically bistable, or does not reflect the
distribution of the pairwise model. This bimodality is first demonstrated
for an experimental dataset comprising 159 neuron activities recorded
from the motor cortex of macaque monkey. Using a reduced maximum-entropy
model, evidence is then provided that this bimodality affects typical
neural recordings of population sizes of a couple of hundreds or more
neurons. As a way to eliminate the bimodality and its ensuing problems,
a modified pairwise model is presented, which -- most important --
has an associated pairwise Glauber dynamics. This model avoids bimodality
thanks to a minimal asymmetric inhibition. It can be interpreted as
a minimum-relative-entropy model with a particular prior, or as a
maximum-entropy model with an additional constraint. The bimodality
problem, the modified maximum-entropy model, and the question of the
relevance of pairwise correlations are presented and discussed from
the general perspective of predicting activity given stimuli, formalized
in simple mathematical terms.}{\footnotesize \par}

{\footnotesize{}}{\footnotesize \par}
\end{abstract}

\section*{{\normalsize{}Author summary}}

{\footnotesize{}}{\footnotesize \par}

{\footnotesize{}Networks of interacting units are ubiquitous in various
fields of biology (gene regulatory networks, neuronal networks, social
structure). If a limited set of observables is accessible, maximum
entropy models provide an unbiased way to construct a statistical
model. The pairwise model only uses the first two moments among those
observables as constraints and therefore yields units that interact
in a pairwise manner. Already at this level, a fundamental problem
arises: if correlations are on average positive, we here show that
the maximum entropy distribution tends to become bimodal. In the application
to neuronal activity, the bimodality is an artefact of the statistical
model. We here explain under which conditions bimodality arises and
present a solution to the problem by introducing a path of collective
negative feedback. This result may point to the existence of a homeostatic
mechanism active in the system that is not part of our set of observable
units.}{\footnotesize \par}

%\ifthenelse{\boolean{isarxiv}}{}{\linenumbers}

\section{Introduction\label{sec:Introduction}}

Understanding the relation between brain activity on one side, and
what we could call the ``state'' of the brain -- the complex combination
of behaviour, stimuli, memory, and thought, which still partly escapes
definition and measurement \citep{Arieli96_1868} -- on the other
side, is a major goal in the study of the brain \citep{kandeletal1991_r2013,lathametal2013}.
We would like to achieve this understanding at a probabilistic level
at least. In very hand-waving terms, we could say it amounts to assigning
probabilities of the form
\begin{equation}
P(\text{brain activity}\cond\text{state}).\label{eq:P_activity_given}
\end{equation}

Given the practically uncountable patterns of activity in the brain
or even in just a small region of it, and the continuous spectrum
and even vagueness of ``states'', assigning such probabilities is
practically impossible and will likely stay that way for the next
few decades. In Bayesian theory we deal with a vague or too large
set of probabilities by introducing one or more statistical models,
which simplify the problem and give it well-defined contours. For
example, we can introduce a set of models $\set{M}$, each of which
includes some multi-dimensional parameter $\bm{\alpha}$, in such
a way that they informationally \emph{screen off} every ``brain activity''
from every ``state'', making them conditionally independent \citep{dawid1979,spohn1980,pearl1988,pearl2000}:
\begin{multline}
P(\text{activity}\cond\bm{\alpha},M,\text{state})=P(\text{activity}\cond\bm{\alpha},M)\\
\text{for all activities and states}.\label{eq:cond_independence_activity_behaviour}
\end{multline}
Then the inverse also holds: $P(\text{state}\cond\bm{\alpha},M,\text{activity})=P(\text{state}\cond\bm{\alpha},M)$.
By the rules of marginalization and total probability we can then
rewrite the probability $P(\text{activity}\cond\text{state})$ as
\begin{equation}
P(\text{activity}\cond\text{state})=\sum_{M}\int P(\text{activity}\cond\bm{\alpha},M)\,P(\bm{\alpha},M\cond\text{state})\,\di\bm{\alpha}.\label{eq:extreme-point-model-activity-behaviour}
\end{equation}

The advantage of this approach appears in the mutual information conditional
on the model $M$ and $\bm{\alpha}$:
\[
I(\text{activity},\text{stimuli}\cond\bm{\alpha},M)=0,
\]
or, paraphrasing Caves \citep{caves2000c}: ``the mutual information
between state and brain activity flows through the model $M$ and
parameters $\bm{\alpha}$''. In this \emph{divide et impera} approach
we deal more easily with $P(\text{activity}\cond\bm{\alpha},M)$ and
$P(\bm{\alpha},M\cond\text{stimuli})$ separately than with the full
probability \prettyref{eq:P_activity_given}, provided the parameter
$\bm{\alpha}$ has much fewer dimensions than the ``activity'' and
``state'' spaces. This parameter then constitutes a coarser but
sufficient description of the activity, or of the state (stimuli,
behaviour, memory, thought processes), or of both. An example of the
first case could be the mean activities and pairwise correlations
of a neuronal population; an example of the second could be the orientation
of a light-intensity gradient on the retina, or ambient temperature.
In the first case, if the model $M$ can be interpreted and motivated
neurobiologically, then it is a ``neural code'' \citep{fersteretal1995,borstetal1999,panzerietal1999,doyaetal2007,quianquirogaetal2013,panzerietal2015}.

The abstract viewpoint just outlined \citep[see][]{freedman1962b,dawid1982,lauritzenetal1984,lauritzen1988,pearl1988,pearl2000,bretthorst2013,dawid2013}
is useful for understanding recent applications of the maximum-entropy
method in neuroscience: the main topic of this paper is in fact a
concrete example of this viewpoint where the model $M$ is a maximum-entropy
model \citep[and references therein]{shannon1948,shannon1949,jaynes1957,jaynes1963,jaynes1967,hobson1969,tribus1969,hobsonetal1973,gulletal1978,jaynes1982b,csiszar1984,meadetal1984,jaynes1985e,csiszar1985,jaynes1986d_r1996,sivia1990,skilling1990,jaynes1994_r2003,georgii2003,csiszaretal2004b},
and the parameter $\bm{\alpha}$ is the empirical means and the pairwise
empirical correlations of the activity of a neuronal population. Does
such a choice of model and parameters give reasonable predictions
$P(\text{activity}\cond\bm{\alpha},M)$? This question has been asked
repeatedly in the neuroscientific literature of the past few years.
Some studies \citep[e.g.,][]{bohteetal2000,schneidmanetal2006,tkaciketal2006,cohenetal2011,granotatedgietal2013}
have tested the suitability of the maximum-entropy distribution --
\prettyref{eq:cond_independence_activity_behaviour} from our perspective
-- for various experimental and simulated ``activities'' and ``states''.
Some studies \citep[e.g.,][]{martignonetal1995,shlensetal2006,mackeetal2009b,barreiroetal2010,ganmoretal2011,shimazakietal2015}
have tested the suitability of pairwise correlations or of higher-order
moments -- our parameter $\bm{\alpha}$. Some studies have done both
at the same time \citep{roudietal2009c,gerwinnetal2010,mackeetal2011b,mackeetal2013}.

Computing the maximum-entropy distribution from moment constraints
-- this is usually called the \emph{inverse problem} -- is very simple
in principle: it amounts to finding the maximum of a convex function
\citep{mead1979,fangetal1997,rockafellar1993}. The maximum can be
searched for with a variety of methods (downhill simplex, direction
set, conjugate gradient, etc. \citep[ch. 10]{pressetal1988_r2007}).
The convex function, however, involves a sum over $\exp(\text{number of neurons})$
terms. For 60 neurons, that is roughly twice the universe's age in
seconds, but modern technologies enable us to record from \emph{hundreds}
of neurons simultaneously \citep{Nicolelis98,Buzsaki04_446,berenyietal2014}.
The convex function must therefore be ``sampled'' rather than calculated,
usually via Markov chain Monte Carlo techniques \citep{james1980,binder1984_r1987,pressetal1988_r2007,mackay1995_r2003,landauetal2000_r2015,andrieuetal2003,krauth2006}.
In neuroscience the Glauber dynamics (also known as Gibbs sampling)
\citep[chap. 29]{glauber1963d,mackay1995_r2003} is usually chosen
as the Markov chain whose stationary probability distribution is
the maximum-entropy one.

\emph{Boltzmann learning }\citep{ackleyetal1985,hintonetal1986_r1999,brodericketal2007}
is such an iterative combination of sampling and maximum search, and
is still considered the most precise method of computing a maximum-entropy
distribution -- at least in the neurosciences, to which our knowledge
on such methods is confined.

A different approach is to approximate the convex function with an
analytic expression, and to find the maximum directly via the study
of the derivatives of this approximation. The mean-field \citep{hartree1928,negeleetal1988_r1998,opperetal2001},
Thouless-Anderson-Palmer \citep{thoulessetal1977,opperetal2001},
and Sessak-Monasson \citep{sessaketal2009,sessak2010} approximations
are examples of this approach, widely used in neuroscience. These
approximations are valid only in limited regions of the domain of
the original convex function, and their goodness is usually checked
against a Boltzmann-learning calculation (as e.g. in \citet{roudietal2009c}).

Outside of Bayesian theory, moment-constrained maximum-entropy models
have also been used in frequentist methods \citep{gruen2009,mackeetal2009}
as generators of surrogate activity data, again via a Glauber dynamics.
Such surrogates are used as ``null hypotheses'' of independence
or pairwise dependence between spike trains \citep{Fujisawa08_823,Gruen03_335,Pipa03a,Pipa03b,Gerstein04}.

The pairwise maximum-entropy model has been used for \emph{experimentally}
recorded activities of populations of a couple hundreds neurons at
most, so far; but its success (or lack thereof) cannot be automatically
extrapolated to larger population sizes. Roudi et al. \citep{roudietal2009b}
gave evidence that the maximized Shannon entropy and other comparative
entropies of such model may present qualitatively different features
above a particular population size. This is possibly also the message
of Tkačik, Mora, et al. \citep{tkaciketal2006,tkaciketal2009,tkaciketal2014b,moraetal2015}
in terms of ``criticality''. (Note that any ``criticality'' in
such models is not a physical property of the population activity,
but of our uncertainty about it. Some choices of models or constraints
may lead to a ``critical'' distribution, other choices may not.)

In the present paper we discuss a feature of the pairwise maximum-entropy
model that may be problematic or undesirable: the marginal distribution
for the population-averaged activity becomes \emph{bimodal}, and one
of the modes may peak at high activities. In other words, maximum-entropy
predicts that the population should fluctuate between a regime with
a small fraction of simultaneously active neurons, and another regime
with a higher fraction of simultaneously active neurons; the fraction
of the second regime be as high as $90\%$. This feature of the maximum-entropy
model seems to have been observed before \citep{bohteetal2000,mackeetal2009b,mackeetal2011},
but never remarked upon.

We also provide evidence that this bimodality is not just a mathematical
quirk: it is bound to appear in applications to populations of more
than a hundred neurons.

The bimodality makes the pairwise maximum-entropy model problematic,
for several reasons.

First, from data reported in the neuroscientific literature \citep[e.g.,][]{cohenetal2011}
the coexistence of two regimes appears neurobiologically unrealistic
-- the more so if the second regime corresponds to $90\%$ of all
units being active.

Second, two complementary problems appear with the Glauber dynamics
and the Boltzmann-learning used to find the model's parameters. If
the minimum between the two probability maxima is shallow, the activity
alternately hovers about either regime for \emph{sustained} periods,
which is again unrealistic, and hence rules out this method to generate
meaningful surrogate data. If the minimum between the two maxima is
deep, the Glauber dynamics becomes practically \emph{non-ergodic,
}and the pairwise model \emph{cannot be calculated at all} via Boltzmann
learning or via the approximations previously mentioned \citep[cf.][\S\ 2.1.3]{landauetal2000_r2015}\citep[chap. 29]{mackay1995_r2003}.
This case is particularly subtle because it can go undetected: the
non-ergodic Boltzmann learning still yields a reasonable-looking distribution,
and this distribution gives back the moments used as constraints when
re-checked with Monte Carlo sampling. However, \emph{this distribution
is not the sought pairwise }maximum-entropy\emph{ distribution}: the
two differ quantitatively also for low activities. This subtle, misleading
self-consistency makes us wonder whether some papers that apply the
model to large populations  are affected by the non-ergodicity,
so that what they find and use is actually not a pairwise maximum-entropy
distribution.

\medskip{}
 The plan of this paper is the following: after some mathematical
and methodological preliminaries we show the appearance of the bimodality
problem with an experimental dataset: the activity of $159$ neurons
recorded from macaque motor cortex. Then we use an analytically tractable
homogeneous pairwise maximum-entropy model (called ``reduced'' model
for reasons explained later) to give evidence that the bimodality
affects larger and larger ranges of datasets as the population size
increases. For example, if the observed Pearson correlations have
a population-average larger than $0.05$, the bimodality is bound
to appear for population sizes of $500$ neurons and above. We show
that experimental datasets of neural-activity are likely to fall within
the bimodality ranges. We also show that the bistability does not
disappear in the inhomogeneous case -- it may become worse.

After analysing the appearance of bimodality and the conditions for
it, we also propose a way to eliminate it: using a slightly modified
pairwise maximum-entropy distribution, which does not suffer from
the bimodality problem. This modified distribution can be interpreted
as arising from the principle of \emph{minimum relative entropy} (also
called minimum discrimination information) \citep[e.g.,][]{kullback1959_r1978,kullbacketal1984,csiszar1985,campbell1985,kullbacketal1951,jaynes1963,renyi1961,hobson1969,chencov1972_t1982,hobsonetal1973,aczeletal1975,gulletal1984,gulletal1984b,kullbacketal1987,kullback1987,gulletal1990_r1999,baezetal2014}
with respect to a neurobiologically motivated reference prior, or
as a maximum-entropy distribution with an additional constraint. The
most important property of this modified distribution is its stationarity
under a modified Glauber dynamics that includes a minimal \emph{asymmetric
inhibition}. In our eyes this gives a neurobiological justification
for using the modified distribution and its Glauber dynamics. We also
show that the modified maximum-entropy distribution is the actual
one obtained via Boltzmann learning or other approximations in the
non-ergodic case -- and thus could be the distribution actually computed
in papers that used such techniques.

We finally bring to a close with a summary about bimodality and its
consequences, a justification of the modified maximum-entropy model
and its reference prior, and on how these models and pairwise correlations
fit within the modelling viewpoint outlined at the beginning of this
section.

Some remarks and extensive references about maximum-entropy methodology
are presented at relevant points in the paper.

Our terminology and mathematical notation conform as much as possible
to ISO (and ANSI, NIST, DIN, JCGM) standards \citep{iso2009,jcgm1993_r2010,jcgm2008,jcgm1993b_r2012,iso2006,iso2006b,nist1994,nist1995},
in the hope of promoting and facilitating interdisciplinary communication.

\section{Results}

\subsection{Preliminaries: maximum-entropy models and Glauber dynamics}

Our study uses three main mathematical objects: the pairwise maximum-entropy
distribution, a ``reduced'' pairwise maximum-entropy distribution,
and the Glauber dynamics associated with them. We review them here
and give some additional remarks and references that we have not seen
elsewhere in the neuroscientific literature. Towards the end of the
paper we will introduce an additional maximum-entropy distribution.

\subsubsection{Pairwise maximum-entropy model\label{sub:Pairwise-maximum-entropy-model}}

First let us make mathematically clear what we mean by ``activity'':
a set of sequences of spikes of $N$ neurons during a finite time
interval $[0,T]$. These spike sequences are discretized: we divide
the time interval into $n$ bins of identical length $\varDelta\text{ equal to }T/n$,
indexed by $t\text{ in }\set{1,\dotsc,n}$. For each neuron $i$,
the existence of one or more spikes in bin $t$ is represented by
$s_{i}(t)=1$, and lack of spikes by $s_{i}(t)=0$. With this binary
representation, the activity of our population at time bin $t$ is
described by a vector: $\yss(t)\coloneqq\bigl(s_{i}(t)\bigr)$. We
will switch freely between vector and component notation for this
and other quantities.

\emph{Time} averages are denoted by a circumflex: $\tav{\dotv}$,
and \emph{population} averages by an overbar: $\av{\dotv}$. The activity
summed over the population at time $t$, or \emph{population-summed
activity}, is denoted by $\yst(t)$, and the \emph{population-averaged
activity} by $\ysm(t)$:
\[
\begin{split}\yst(t) & \coloneqq\sum_{i=1}^{N}s_{i}(t)\in\set{0,1,\dotsc,N},\\
\ysm(t) & \coloneqq\frac{1}{N}\sum_{i=1}^{N}s_{i}(t)\equiv\frac{1}{N}S(t)\in\set{0,1/N,\dotsc,1}.
\end{split}
\]
The time-averaged activity of neuron $i$ is denoted by $m_{i}$:
\begin{equation}
m_{i}\coloneqq\tav{s_{i}(t)}\coloneqq\frac{1}{T}\sum_{t=1}^{n}s_{i}(t),\label{eq:m_i}
\end{equation}
and the time average of the product of the activities of the neuron
pair $ij$, called \emph{coupled activity}, is denoted by $g_{ij}$:
\begin{equation}
g_{ij}\coloneqq\tav{s_{i}(t)\,s_{j}(t)}\coloneqq\frac{1}{T}\sum_{t=1}^{n}s_{i}(t)\,s_{j}(t).\label{eq:g_ij}
\end{equation}
These time averages are used as constraints for the maximum-entropy
model, as presently explained.

The \emph{pairwise }maximum-entropy\emph{ }statistical model \citep{martignonetal1995,bohteetal2000,Schneidman06_1007,Shlens06_8254}
assigns a time-independent probability distribution for the population
activity $\yss(t)$ of the form (time is therefore omitted in the
notation):
\begin{equation}
\begin{split}\PP(\yss\cond\hh,\jj) & =\frac{1}{\zp(\hh,\jj)}\exp\bigl(\sum_{i}\mu_{i}s_{i}+\sum_{i<j}\varLambda_{ij}s_{i}s_{j}\bigr),\\
\zp(\hh,\jj) & \coloneqq\sum_{\yss}\exp\bigl(\sum_{i}\mu_{i}s_{i}+\sum_{i<j}\varLambda_{ij}s_{i}s_{j}\bigr);
\end{split}
\label{eq:pairwise-maxent}
\end{equation}
the Lagrange multipliers $\hh(\mm,\ygg)$ and $\jj(\mm,\ygg)$ are
determined by enforcing the equality of the time averages \prettyref{eq:m_i}
and \prettyref{eq:g_ij} with the single- and coupled-activity expectations,
with their usual definitions
\begin{equation}
\expep{s_{i}}\coloneqq\sum_{\yss}s_{i}\,\PP(\yss),\qquad\expep{s_{i}s_{j}}\coloneqq\sum_{\yss}s_{i}s_{j}\,\PP(\yss)\label{eq:definition_expectations}
\end{equation}
(or in matrix form $\expep{\yss}\coloneqq\sum_{\yss}\yss\,\PP(\yss)$
and $\expep{\yss\yss\T}\coloneqq\sum_{\yss}\yss\yss\T\,\PP(\yss)$):
\begin{equation}
\expep{s_{i}}=m_{i}\quad\text{and}\quad\expep{s_{i}s_{j}}=g_{ij}.\label{eq:constraints_single_and_couple}
\end{equation}
Noting that $\expep{s_{i}}=\PP(s_{i}=1)$, and $\expep{s_{i}s_{j}}=\PP(s_{i}=1,s_{j}=1)$,
we see that the constraints above are equivalent to fully fixing the
single-neuron marginal probabilities of $\PP$ and partly fixing its
two-neurons marginal probabilities.

If we introduce the covariances $\cc$ and Pearson correlation coefficients
$\rrho$,
\begin{equation}
\begin{split}c_{ij} & \coloneqq\expe{s_{i}s_{j}}-\expe{s_{i}}\expe{s_{j}},\\
\rho_{ij} & \coloneqq\frac{c_{ij}}{\sqrt{[\expep{s_{i}^{2}}-\expep{s_{i}}^{2}]\,[\expep{s_{j}^{2}}-\expep{s_{j}}^{2}]}},
\end{split}
\end{equation}
the constraints above are jointly equivalent to
\begin{equation}
\expep{s_{i}}=m_{i}\quad\text{and}\quad c_{ij}=g_{ij}-m_{i}m_{j}\label{eq:constraints_single_covariances}
\end{equation}
or
\begin{equation}
\expep{s_{i}}=m_{i}\quad\text{and}\quad\rho_{ij}=\frac{g_{ij}-m_{i}m_{j}}{\sqrt{(m_{i}-m_{i}^{2})\,(m_{j}-m_{j}^{2})}}\label{eq:constraints_single_pearsoncorr}
\end{equation}
Note that the covariance constraints $c_{ij}=g_{ij}-m_{i}m_{j}$
by themselves are not convex, i.e., they do not define a convex subset
in the probability simplex on which the entropy is maximized. The
Lagrange-multiplier method does not guarantee the uniqueness of the
solution if \emph{only} the covariances are constrained. Uniqueness
has to be checked separately \citep{csiszar1984,meadetal1984,garrett1990b,fangetal1997,csiszaretal2004b}.
On the other hand, the constraints $\expep{s_{i}}=m_{i}$ and $\expep{s_{i}s_{j}}=g_{ij}$
are separately convex, thus their conjunction $\expep{s_{i}}=m_{i}\Land\expep{s_{i}s_{j}}=g_{ij}$
is convex too, and the bijective correspondence of the latter with
$\expep{s_{i}}=m_{i}\Land c_{ij}=g_{ij}-m_{i}m_{j}$ guarantees that
the latter set of constraints is convex as well. What we have said
about the covariances $\cc$ also holds for the correlations $\rrho$.

It is important to remember that $(\mm,\ygg)$ are physically measurable
quantities, independent of the observer, whereas $\bigl(\expep{s_{i}},\expep{s_{i}s_{j}}\bigr)$
depend on the observer's uncertainty, quantified by her probability
assignment, and are not physically measurable. Therefore the constraints
\prettyref{eq:constraints_single_and_couple} are not trivial definitions
(``$\coloneqq$'') or equivalences (``$\equiv$''). In fact, in
particular situations it does not make sense to enforce some of the
constraints \citep{portamana2009}; this also depends on \emph{what
is our uncertainty about}. Let us explain this point.

The maximum-entropy distribution represents our uncertainty about
the population activity for each of the \emph{given} time bins, $t\in\set{1,\dotsc,n}$;
this can be shown by symmetry and combinatorial arguments within the
probability calculus \citep{gulletal1978,frieden1980,jaynes1982b,csiszar1984,jaynes1985e,csiszar1985,jaynes1986d,jaynes1986d_r1996,skilling1990,csiszaretal2004b,portamana2009}.
Sometimes the maximum-entropy distribution is also used to represent
someone's uncertainty about a \emph{new} observation about a new time
bin, e.g. $t\text{ equal to }n+1$. But such use implies additional
assumptions and a particular prior that are not always justified \citep{portamana2009}.
Here is an example: suppose the time average of the coupled activity
of neurons $1$ and $2$ vanishes: $g_{1\,2}\coloneqq\frac{1}{T}\sum_{t=1}^{n}s_{1}(t)\,s_{2}(t)=0$
(this happens for a couple of pairs in our data). If we enforce the
constraint $\expep{s_{1}s_{2}}=g_{1\,2}=0$, then maximum-entropy
says that it is \emph{impossible} that neurons $1$ and $2$ spike
together: $\PP(s_{1}=1,s_{2}=1)=0$ (the corresponding Lagrange multiplier
$\varLambda_{1\,2}=-\infty$). This prediction makes sense if we are
speaking about any of our $n$ time bins -- in fact, $g_{1\,2}=0$
means that neurons $1$ and $2$ have never spiked together in our
data, so the prediction is right. But it is an unreasonable prediction
about a future or past time bin that is not part of our data: just
because neurons $1$ and $2$ have not spiked simultaneously in our
$n$ data bins, we cannot conclude that it is \emph{impossible} for
them to spike or have spiked simultaneously in the future ($t>n$)
or in the past ($t<0$). Therefore, when some constraints assume extreme
values, as $g_{1\,2}=0$ in our example, it is not meaningful to use
the maximum-entropy model for \emph{new} predictions outside the given
dataset. In this case it is more appropriate to use the full (Bayesian)
probability calculus \citep{berger1980_r1985,bernardoetal1994,mackay1995_r2003,gelmanetal1995_r2004,zellner1971_r1996,portamana2009,westetal1989_r1997},
possibly with maximum-entropy ideas on a more abstract level (space
of prior distributions) \citep{skilling1989b,rodriguez1991,caticha2001,rodriguez2002,rodriguez2003,catichaetal2004}.

Contrary to what is sometimes stated in the literature, it is not
true that the maximum-entropy model can only be used if the time sequence
of activities is ``stationary''. This model represents a guess about
the activities in the sequence, given \emph{time-average} information.
This guess, therefore, has to be time-invariant by symmetry: any time-dependent
information has been erased by the time averaging. In other words,
it is our guess which is ``stationary'', not the physical data;
but it is still a good guess, given the time-independent information
provided. With time-dependent constraints we would obtain a time-dependent
maximum-entropy distribution \citep[cf.][]{hertzetal2011_r2013}:
this application of the maximum-entropy principle is called ``maximum-calibre''
\citep{jaynes1979b,jaynes1980c,jaynes1985b,grandy2008,leeetal2012,presseetal2013}.

\subsubsection{Reduced maximum-entropy model\label{sub:Reduced-model}}

If the time-averaged activities $\mm$ are homogeneous, i.e. equal
to one another and to their population average $\ymmm$, and the $N\,(N-1)/2$
time-averaged coupled activities $\ygg$ are also homogeneous with
population average $\yggm$, $\yggm\coloneqq\tfrac{2}{N\,(N-1)}\sum_{i<j}g_{ij}$,
then the pairwise maximum-entropy distribution has homogeneous Lagrange
multipliers by symmetry: $\mu_{i}=\yhhr$ and $\varLambda_{ij}=\yjjr$.
It reduces to the simpler and analytically tractable form

\begin{gather}
\begin{split}\PR(\yss\cond\yhhr,\yjjr) & =\frac{1}{\zr(\yhhr,\yjjr)}\exp[\yhhr N\ysm+\tfrac{1}{2}\yjjr\,N\ysm\,(N\ysm-1)],\\
\zr(\yhhr,\yjjr) & \coloneqq\sum_{\yss}\exp[\yhhr N\ysm+\tfrac{1}{2}\yjjr\,N\ysm\,(N\ysm-1)],
\end{split}
\label{eq:pairwise-reduced-maxent}
\end{gather}
which assigns equal probabilities to all those activities $\yss$
that have the same population average $\ysm$. In this homogeneous
case, the values of the multipliers are equal to their averages: $\mu_{i}=\yhhr=\yhhm\coloneqq\frac{1}{N}\sum_{i}\mu_{i}$
and $\varLambda_{ij}=\yjjr=\yjjm\coloneqq\frac{2}{N\,(N-1)}\sum_{i<j}\varLambda_{ij}$. 

This simpler distribution could be interpreted as an approximation
of the pairwise maximum-entropy one, achieved by disregarding the
inhomogeneities. But it is also an exact maximum-entropy distribution
in its own right, obtained by only constraining the expectations for
the \emph{population sums} of the single and coupled activities,
\[
\sum_{i}s_{i}=S=N\ysm,\qquad\sum_{i<j}s_{i}s_{j}=S\,(S-1)/2=N\ysm\,(N\ysm-1)/2,
\]
to be equal to their measured time averages:
\begin{equation}
\exper{N\ysm}=N\ymmm\quad\text{and}\quad\expebr{N\ysm\,(N\ysm-1)}=\frac{N\,(N-1)}{2}\yggm\coloneqq\sum_{i<j}g_{ij}\label{eq:constraints_single_and_couple-reduced}
\end{equation}
(or equivalently constraining the population averages.)

For this reason we call the model \prettyref{eq:pairwise-reduced-maxent}
a \emph{reduced} (pairwise) maximum-entropy model. If the time-averages
are homogeneous, then $\yhhr=\yhhm=\mu_{i}$, $\yjjr=\yjjm=\varLambda_{ij}$
and the reduced and full pairwise model coincide. But in the inhomogeneous
case the multipliers of the reduced model are \emph{not} equal to
the averages of the pairwise one: $\yhhr\ne\yhhm$, $\yjjr\ne\yjjm$.

It is straightforward to derive the probability distribution for the
population average $\ysm$ in this model, owing to its symmetry: if
the average is $\ysm$, there must be $N\ysm$ active neurons in the
population, and there are $\tbinom{N}{N\ysm}$ ways in which this
is possible, all having equal probability given by \prettyref{eq:pairwise-reduced-maxent}.
Therefore,
\begin{equation}
\begin{split}\PR(\ysm\cond\yhhr,\yjjr) & =\frac{1}{\zr(\yhhr,\yjjr)}\binom{N}{N\ysm}\,\exp[\yhhr N\ysm+\tfrac{1}{2}\yjjr\,N\ysm\,(N\ysm-1)],\\
\zr(\yhhr,\yjjr) & \coloneqq\sum_{\ysm}\binom{N}{N\ysm}\,\exp[\yhhr N\ysm+\tfrac{1}{2}\yjjr\,N\ysm\,(N\ysm-1)].
\end{split}
\label{eq:probability-of-average}
\end{equation}
This probability distribution $\PR(\ysm)$ can, in turn, also be obtained
applying a minimum-relative-entropy principle \citep{kullback1959_r1978,kullbacketal1984,csiszar1985,campbell1985,kullbacketal1951,jaynes1963,renyi1961,chencov1972_t1982,aczeletal1975,gulletal1984,gulletal1984b,kullbacketal1987,kullback1987,gulletal1990_r1999},
i.e. minimizing the relative entropy (or discrimination information)
\begin{equation}
H(P,P_{0})\coloneqq\sum_{\ysm}P(\ysm)\ln\frac{P(\ysm)}{P_{0}(\ysm)}
\end{equation}
 of $P(\ysm)$ with respect to the reference distribution $P_{0}(\ysm)=2^{-N}\,\tbinom{N}{N\ysm}$
while constraining the first two moments of $\PR(S)$, or equivalently
its first two factorial moments \citep{potts1953}, $\bigl(\expe S,\expe{S\,(S-1)/2}\bigr)$.

It is easy to see that in this model, by symmetry, we also have
\begin{gather}
\exper{s_{i}}=\exper{\ysm},\qquad\exper{s_{i}s_{j}}=\expebr{\frac{N\ysm\,(N\ysm-1)}{N\,(N-1)}},\\
c_{ij}=\yccm=\expebr{\frac{N\ysm\,(N\ysm-1)}{N\,(N-1)}}-\exper{\ysm}^{2},\qquad\rho_{ij}=\av{\rho}=\frac{\yccm}{\exper{\ysm}-\exper{\ysm}^{2}},\label{eq:correlation_pearson_pop-averages}
\end{gather}
and $\Bigl(\exper{\ysm},\expebr{\frac{N\ysm\,(N\ysm-1)}{N\,(N-1)}}\Bigr)$,
$\bigl(\exper{\ysm},\yccm\bigr)$, $\bigl(\exper{\ysm},\av{\rho}\bigr)$
are equivalent sets of constraints (but $\yccm$ and $\yrhom$ by
themselves are not convex).

The reduced maximum-entropy model is mathematically very convenient,
because the Lagrange multipliers $\yhhr,\yjjr$ can be easily found
numerically (with standard convex-optimization methods like downhill
simplex, direction set, conjugate gradient, etc. \citep[ch. 10]{pressetal1988_r2007})
with high precision even for large (e.g., thousands) population sizes
$N$.

Summarizing: the reduced maximum-entropy model can be seen as: 1.
the form taken by the pairwise maximum-entropy model in the case of
homogeneous single and couple activities; 2. an approximation to the
pairwise maximum-entropy model in the case of inhomogeneous single
and couple activities; 3. a maximum-entropy model in its own right,
that uses less information than the full pairwise model.

\subsubsection{Glauber dynamics}

The maximum-entropy distributions above do not make any prediction
about the dynamical or kinematical properties of the population activity,
like first-passage times. They are, however, identical in form to
the stationary distribution of an asynchronous Glauber dynamics \citep{glauber1963d}
with symmetric ``couplings'' $\jj$, sometimes interpreted as symmetric
synaptic couplings, and ``biases'' $\hh$, sometimes interpreted
as either a threshold or external input controlling the base activity
of individual neurons. In the reduced maximum-entropy model these
parameters are homogeneous: $\varLambda_{ij}=\yjjr$, $\mu_{i}=\yhhr$.
The full and reduced maximum-entropy distributions give some information
about this particular dynamics, like the appearance of metastable
or most probable population-average states.

If we assume that our uncertainty about the \emph{evolution} of the
population activity can be modelled by the Glauber dynamics of a binary
network, we can choose the $\hh,\jj$ parameters determined by the
constraints \prettyref{eq:constraints_single_and_couple} and thus
generate surrogate data that -- if the dynamics is ergodic -- will
have infinite-time-average activities as our initial experimentally
observed data.

The formulae of the pairwise maximum-entropy model are similar or
even identical to the formulae of the Lenz-Ising or Sherrington-Kirkpatrick
spin model \citep{lenz1920,ising1925,peierls1935,onsager1944,kawasaki1972,sherringtonetal1975,kirkpatricketal1978}.
This similarity is useful: it allows us to borrow some mathematical
techniques, approximations, and intuitive pictures developed for one
model, and to apply them to the other. Yet we purposely emphasize
the probability-calculus viewpoint and avoid any ``explanation''
via statistical-mechanical analogies and their related concepts and
jargon. On the whole, such analogies are conceptually limitative and
pedagogically detrimental because they put the logical cart before
the logical horse: the logical route is not \emph{statistical mechanics
→ maximum-entropy}, but \emph{probability calculus + physics → maximum-entropy
→ statistical mechanics} \citep[chaps I--IV]{gibbs1902}\citep{emchetal2002,grandy2008,balian1982_t2007,balian1982b_t2007,grandy1987,grandy1988,hobson1966,jaynes1967,hobsonetal1968,hobson1971,jaynes1979b,jaynes1985b,maesetal2002_r2003}
\citep[see also][]{maxwell1871b,maxwell1875,maxwell1878,einstein1902}.
There are in fact important differences between the two models and
their quantities. In no particular order:

First: in the case of the Lenz-Ising model, the microscopic state
is unknown: we try to guess it from macroscopic properties; this is
a problem of inference \emph{within} a model (the energy-constrained
maximum-entropy model itself is not brought into question). The opposite
holds for the pairwise maximum-entropy model: the ``microscopic state''
(activity) is known, and we try to find the ``macroscopic properties''
that lead to a good guess about it; this is a problem of inference
\emph{of} a model \citep[cf.][]{mackay1992}.

Second: the Lenz-Ising model has one macroscopic quantity as constraint:
the total energy, which has one associated Lagrange multiplier: the
statistical temperature. The pairwise maximum-entropy model has $N+(N^{2}-N)/2$
constraints, with as many associated Lagrange multipliers. The difference
of constraints between the two models implies essential differences
between their entropies; negligence of such differences leads to variants
of the Gibbs paradox \citep{grad1952,grad1961,grad1967,jaynes1992,lebowitzetal2003}.

Third: The couplings and external fields that appear in the energy
of the Lenz-Ising model are measurable physical quantities. The mathematically
similar Lagrange multipliers of the pairwise maximum-entropy model
are statistical parameters and cannot be measured -- they encode our
ignorance. In particular, the expression ``$\sum_{i}\mu_{i}s_{i}+\sum_{i<j}\varLambda_{ij}s_{i}s_{j}$''
-- what Gibbs \citep{gibbs1902} calls \emph{index of probability}
-- is \emph{not} an energy. The following exercise shows why: Assume
$\mu_{i}=-3$, $\varLambda_{ij}=0.04$, consider a transition from
a state $S=7$ to the state $S=0$, and calculate by how many metres
we could lift a $1\;\mathrm{kg}$ weight if that ``energy'' difference
could be converted into mechanical energy.

These differences do not stop us from using mathematical techniques
common to the two models to our advantage.

\subsection{The problem: bimodality, bistability, non-ergodicity\label{sub:The-problem-bimodality}}

We first show how the bistability problem subtly appears with a set
of experimental data, then explore its significance for larger population
sizes.

\subsubsection{Example with experimental data}

Our data consists in the activity of a population of $159$ neurons
($N=159$) from motor cortex of macaque monkey, recorded for $15$
minutes using a 100-electrode ``Utah'' array. See \citep{Riehle13_48}
for the experimental setup. The monkey was in a so-called ``state
of ongoing activity” \citep{Arieli96_1868}, i.e. sitting on a chair
without performing any task.

\begin{figure}[h]
\includegraphics[width=1\textwidth]{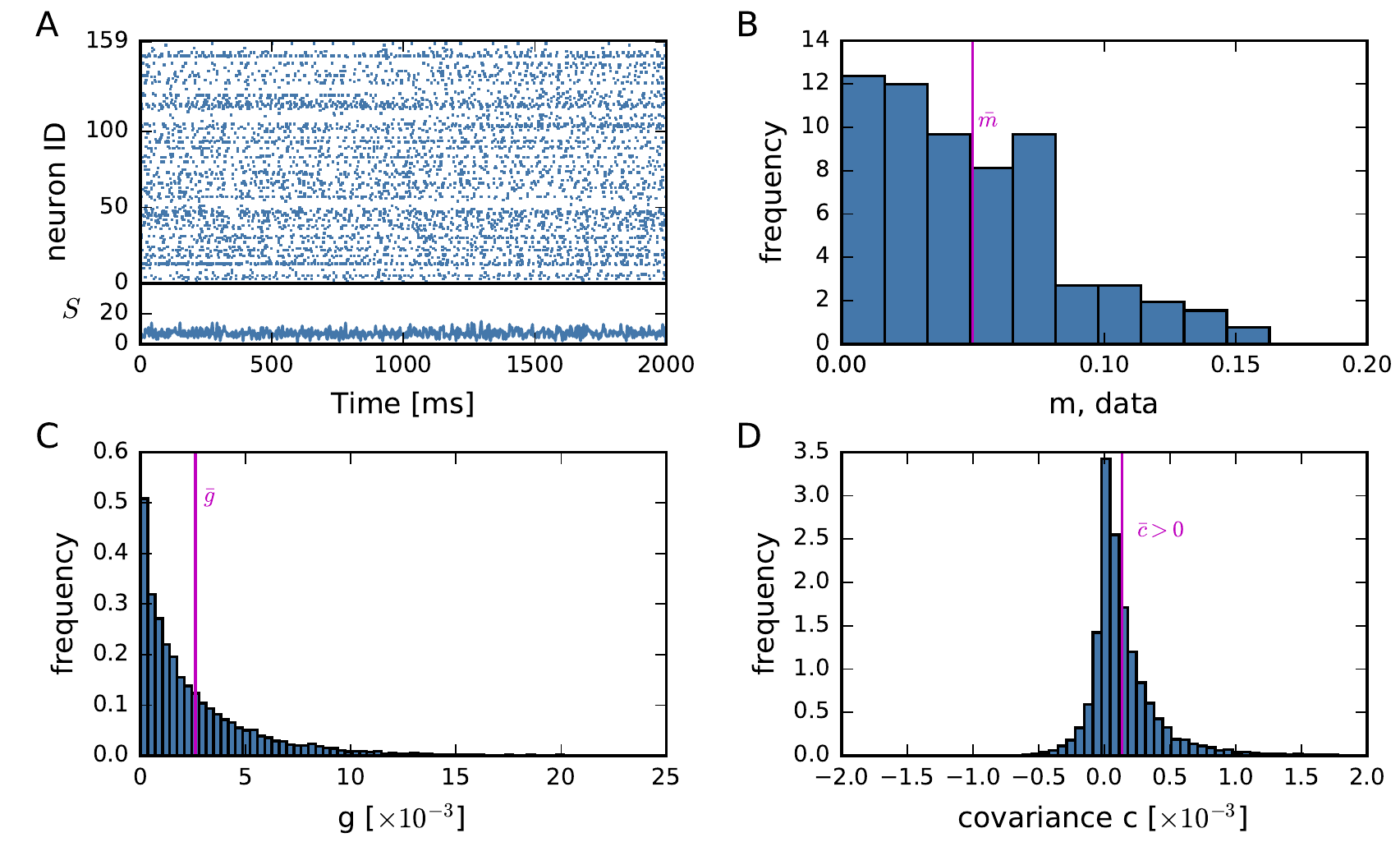}

\caption{\textbf{Experimental data and their empirical first- and second-order
statistics.} (\textbf{A}) Dot display of $159$ parallel spike recordings
of macaque monkey during a state of ``ongoing activity''. The experimental
data are recorded with a 100-electrode ``Utah'' array (Blackrock
Microsystems, Salt Lake City, UT, USA) with 400 \textmu m interelectrode
distance, covering an area of 4\texttimes 4 mm (subsession: s131214-002).
The population-summed activity $S(t)$ is the sum of the number of
active neurons within time bin $t$. The time bins have width $\Delta=3\protect\ms$.
(\textbf{B}) Population distribution of the time-averaged activities
$m_{i}$, \prettyref{eq:m_i}. The vertical line marks the population
average, $\bar{m}\protect\coloneqq\frac{1}{N}\sum_{i}m_{i}$. (\textbf{C})
Population distribution of the time-averaged coupled activities $g_{ij}$,
\prettyref{eq:g_ij}. The vertical line marks the population average,
$\protect\yggm\protect\coloneqq\tfrac{2}{N\,(N-1)}\sum_{i<j}g_{ij}$.
(\textbf{D}) Population distribution of the covariances $c_{ij}=g_{ij}-m_{i}m_{j}$.
The vertical line again marks the population average, $\protect\yccm\protect\coloneqq\tfrac{2}{N\,(N-1)}\sum_{i<j}c_{ij}$;
we have positive average correlations, $\bar{c}>0$. (Histograms bins
in B, C, D computed with Knuth's rule \citep{knuth2006_r2013}). Data
courtesy of A. Riehle and T. Brochier.}
\label{fig:experimental_data}
\end{figure}

Figure \ref{fig:experimental_data}A shows a two-second raster plot
of the activity $\yss(t)$ of the recorded neurons. The time-varying
population-summed activity $S(t)$ is shown underneath. The time-averaged
single and coupled activities $m_{i},\,g_{ij}$, and corresponding
empirical covariances $c_{ij}$ from the data are shown in panels
B, C, D. The population averages of these quantities are
\begin{equation}
\ymmm\approx0.0499,\quad\yggm\approx0.00261,\quad\yccm\approx0.000135,\quad\av{\rho}\approx0.00319.\label{eq:pop-averaged-constraints_numerical_values}
\end{equation}

\begin{figure}[h]
\includegraphics[width=1\textwidth]{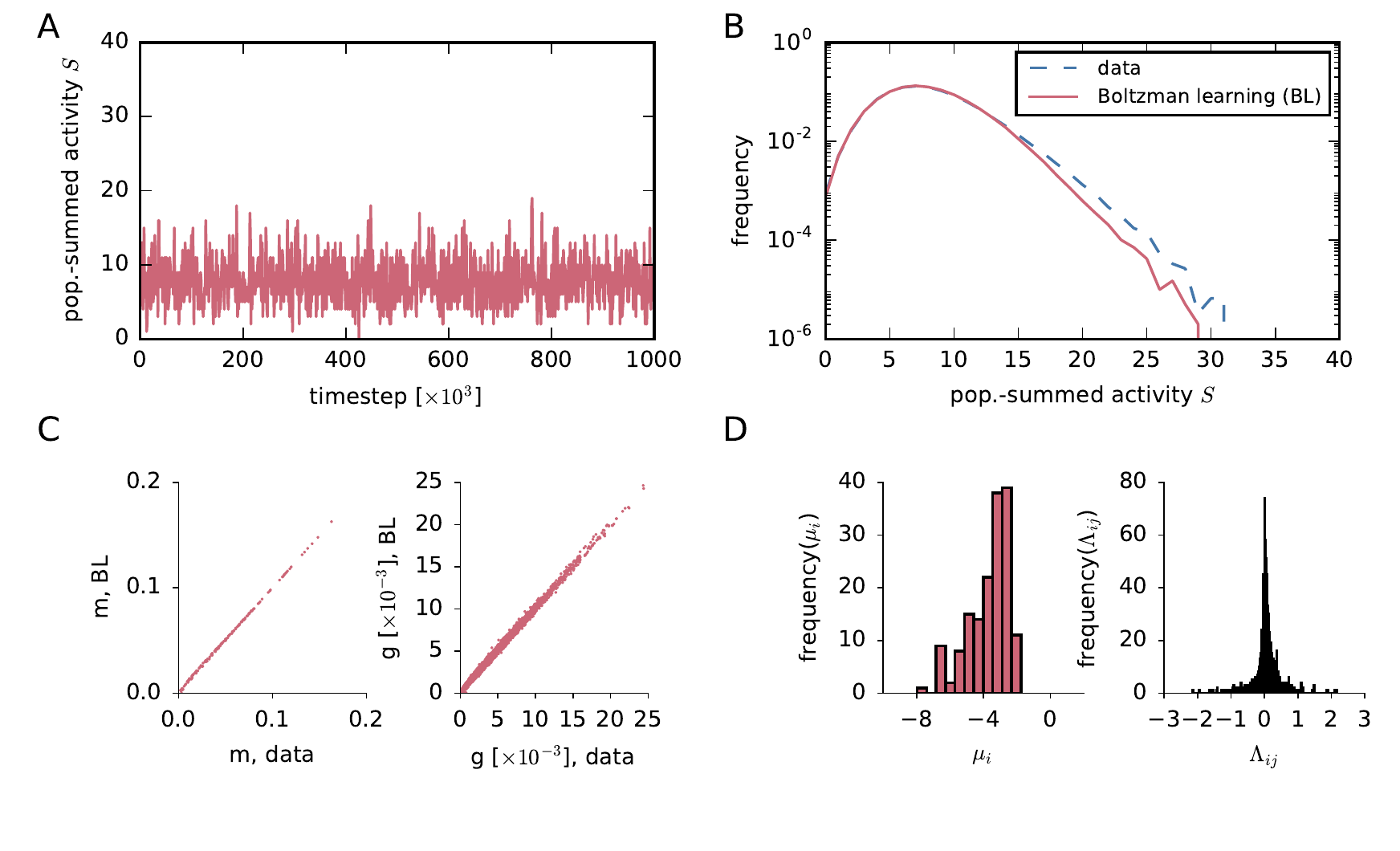}\caption{\textbf{Results of Boltzmann learning}. (\textbf{A}) Population-summed
activity $S(t)$ of $N=159$ neurons, obtained via Glauber dynamics
in $10^{6}$ timesteps. The couplings $\varLambda_{ij}$ and biases
$\mu_{i}$ of the Glauber dynamics are the Lagrange multipliers, shown
in panel D, found by Boltzmann learning from the experimental time-averages
$m_{i}$ and $g_{ij}$ of \prettyref{fig:experimental_data}. (\textbf{B})
Red, solid: Probability distribution of the population-summed activity,
sampled via the Glauber dynamics of panel A. Blue, dashed: empirical
distribution of the population-summed activity from our dataset. (\textbf{C})
Time averages $m_{i}$ and $g_{ij}$ obtained from Boltzmann learning,
versus experimental ones. (\textbf{D}) Population distribution of
the Lagrange multipliers $\mu_{i}$ and $\varLambda_{ij}$ obtained
via Boltzmann learning. (Histogram bins in D computed with Knuth's
rule \citep{knuth2006_r2013}).\label{fig:first_Boltzmann}}
\end{figure}

Let us find, via Boltzmann learning, the Lagrange multipliers of
the pairwise maximum-entropy model constrained by the single and coupled
activities plotted in \prettyref{fig:experimental_data}B--C. At each
iteration, the sampling phase of the Boltzmann learning has $10^{6}$
timesteps; an example is shown in \prettyref{fig:first_Boltzmann}A.
Note that the number of timesteps exceeds the ones used in Roudi et
al. \citep{roudietal2009c} ($N=200$) by a factor of ten and that
in Broderick et al. \citep{brodericketal2007} ($N=40$) by a factor
of three. The learning converges and we obtain the Lagrange multipliers
$(\mu_{i},\varLambda_{ij})$ whose distributions are shown in \prettyref{fig:first_Boltzmann}D.
The final single and coupled activities are shown in \prettyref{fig:first_Boltzmann}C:
they appear very close to the experimental ones. Sampling once more
the maximum-entropy distribution with the obtained Lagrange multipliers,
we obtain the population-average probability distribution, shown in
\prettyref{fig:first_Boltzmann}B against the empirical one. The
tails of the two distributions differ, but this does not concern us
now.

The results of the Boltzmann learning do not show any inconsistency
at this point.

\begin{figure}[h]
\includegraphics[width=1\textwidth]{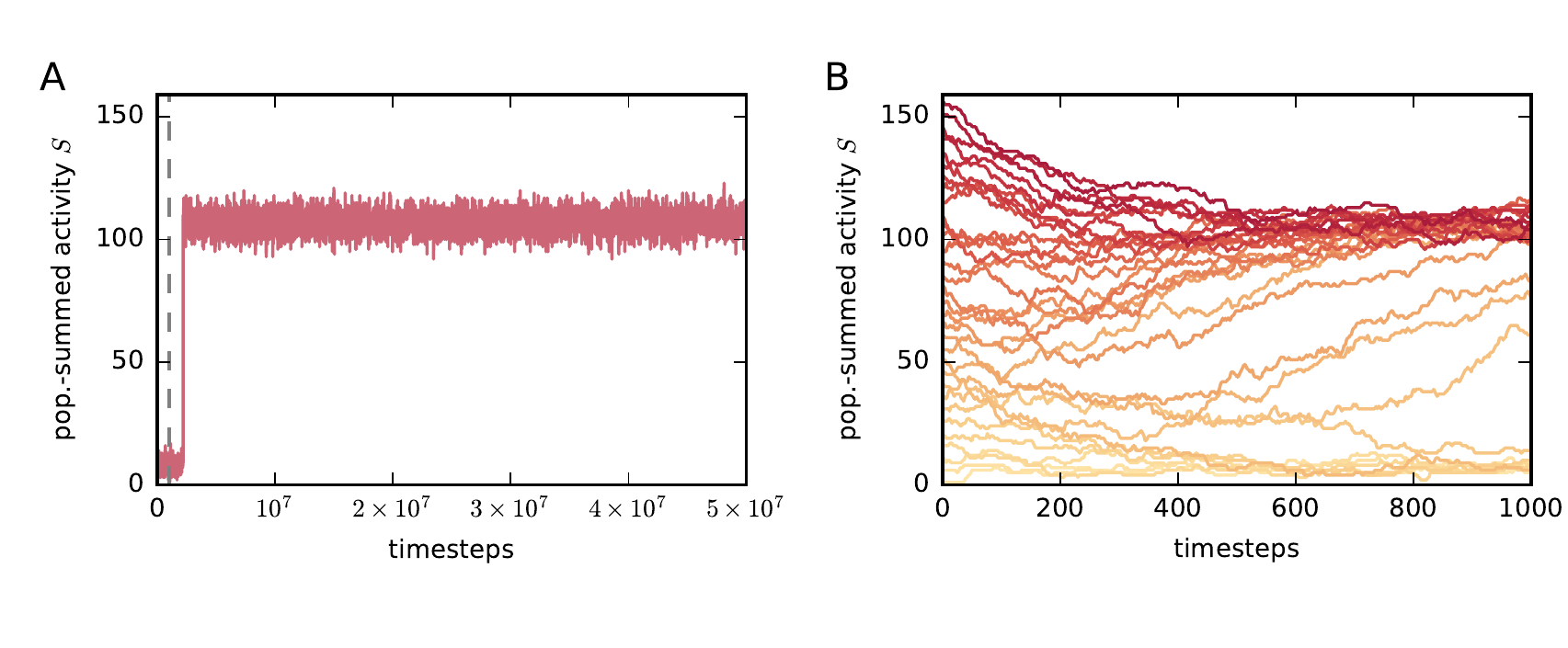}\caption{\textbf{Longer sampling: bistability.} (\textbf{A}) Population-summed
activity $S(t)$ obtained via Glauber dynamics, as in \prettyref{fig:first_Boltzmann}A,
but with longer sampling: $5\times10^{7}$ timesteps. The dashed grey
line marks the end of the previous sampling of \prettyref{fig:first_Boltzmann}A.
(\textbf{B}) Population-summed activities $S(t)$ obtained from several
instances of Glauber dynamics. Each instance starts with a different
initial population activity $\protect\yss(0)$, having different initial
population sum $S(0)$, and is represented by a different red shade,
from $S(0)=0$ (light red) to $S(0)=N$ (dark red).\label{fig:jump_longer_sampling}}
\end{figure}

But now we sample the distribution for a much longer time, say $5\times10^{7}$
steps. \prettyref{fig:jump_longer_sampling}A shows what happens in
a real instance. After roughly $2\times10^{6}$ steps, the population
jumps to a high-activity regime and remains there till the end of
the sampling. We have discovered that the Glauber dynamics has an
additional metastable high-activity regime. How many metastable regimes
could there be? Starting the dynamics from states having different
population-averages, we see that there are two metastable regimes;
see \prettyref{fig:jump_longer_sampling}B. This means that the actual
distribution associated with the Lagrange multipliers of \prettyref{fig:first_Boltzmann}D
must be \emph{bimodal.}

 The discovery of the second metastable regime has important implications
and causes quite a few problems: 
\begin{itemize}
\item Our Boltzmann learning had actually not yet converged: if we sample
long enough to allow the exploration of both metastable regimes, we
find different time-averages of the single and coupled activities
from those of \prettyref{fig:first_Boltzmann}C, in complete disagreement
with the experimental ones.
\item The Lagrange multipliers we obtained, \prettyref{fig:first_Boltzmann}D,
are therefore not correct. Hence \emph{the probability distribution
obtained from the initial Boltzmann learning is not the true pairwise
maximum-entropy distribution.}
\item In order to sample the probability distribution around both modes
and estimate their relative heights, we would need to observe many
jumps between the two metastable regimes. The time required to observe
one such jump seems to be larger than $5\times10^{7}$ timesteps (we
did not wait for longer), which is impractically long. For practical
purposes the Glauber dynamics is \emph{non-ergodic}, and the Boltzmann
learning cannot proceed: \emph{we cannot find the true pairwise distribution}
within reasonable times.
\item The Sessak-Monasson approximation \citep{sessaketal2009,sessak2010}
is not correct either, because it gives a solution very close to the
erroneous Boltzmann-learning one; the Lagrange multipliers of the
sought-for maximum-entropy distribution evidently lie outside of its
radius of convergence.
\end{itemize}
The reason why the initial result seemed self-consistent is that the
sampling phase was too brief compared to the time needed to explore
the full distribution: the latter time is so long that the dynamics
is \emph{non-ergodic} for computational purposes. This non-ergodicity
effectively truncates the sampling at states $\yss$ for which $\ysm\lesssim\theta$,
where $\theta$ is the population-averaged activity at the trough
between the two metastable regimes. In other words, the Lagrange multipliers
$\hh,\jj$ that we found belong to the ``truncated'' distribution
\begin{equation}
\PT(\yss\cond\hh,\jj,\theta)\propto\begin{cases}
\exp\bigl(\sum_{i}\mu_{i}s_{i}+\sum_{i<j}\varLambda_{ij}s_{i}s_{j}\bigr), & \ysm\leqslant\theta,\\
0, & \ysm>\theta,
\end{cases}\label{eq:effective_truncated_P}
\end{equation}
which is \emph{not} a pairwise maximum-entropy distribution. The expectations
of the single and coupled activities with respect to this distribution
equal the experimental time averages:
\begin{equation}
\expet{s_{i}}=m_{i},\qquad\expet{s_{i}s_{j}}=g_{ij},
\end{equation}
but, again, these expectations do \emph{not} come from the true pairwise
maximum-entropy distribution, whose Lagrange multipliers remain unknown. 

Everything is self-consistent as long as we use the truncated distribution
$\PT$, but this is not the pairwise one $\PP$. This remark will
be important later on.

Now the question is whether the correct, sought-for maximum-entropy
distribution is also bimodal, or the Boltzmann learning simply encountered
a bimodal distribution during its search of the correct one in the
space of probabilities.

We make an educated guess by examining the analytically tractable
reduced maximum-entropy model $\PR$, \prettyref{eq:pairwise-reduced-maxent}.
Using the population-averaged single and coupled activities as constraints,
$\exper{\av{s_{i}}}=\ymmm$ and $\exper{\av{s_{i}s_{j}}}=\yggm$ from
\prettyref{eq:pop-averaged-constraints_numerical_values}, we numerically
find the Lagrange multipliers of the reduced model:
\begin{equation}
\yhhr\approx-3.259,\qquad\yjjr\approx0.03859.\label{eq:Lagrange_values_reduced}
\end{equation}
Note that in this case there is no sampling involved -- the distribution
can be calculated analytically -- so the values above are correct
within the numerical precision of the maximization procedure (interior-point
method \citep[chap. 10]{pressetal1988_r2007}). The values of the
expected single and couple activities, re-obtained by explicit summation
(not sampling) from the corresponding reduced maximum-entropy distribution,
agree with the values \prettyref{eq:pop-averaged-constraints_numerical_values}
to seven significant figures.

The resulting reduced maximum-entropy distribution for the population-summed
activity, $\PR(S\cond\yhhr,\yjjr)$, is shown in \prettyref{fig:Appearance-of-bistable}A,
together with the experimental time-frequency distribution of our
data. It shows a second maximum at roughly $90\%$ activity. An exact
analysis of small-population cases, and an analysis of large-population
cases with a maximum-entropy model constrained by the population variance
of the second moments, corresponding to constraining $\expeb S$,
$\expe{S^{2}}$, $\expeb{\tfrac{S\,(S-1)}{N\,(N-1)}-\biggl(\tfrac{S\,(S-1)}{N\,(N-1)}\biggr)^{2}}$
(neither analysis is discussed here), show that if a reduced maximum-entropy
model is bimodal, the full inhomogeneous model is also bimodal, with
a heightened second mode shifted towards lower activities with respect
to the reduced model.

We therefore expect the correct, full pairwise maximum-entropy distribution
for our data to be bimodal.

We will shortly propose a solution to the eliminate the bimodality.
The basic idea behind this solution is easily grasped by first presenting
an intuitive picture of how the bimodality arises.

\subsubsection{Intuitive understanding of the bimodality: Glauber dynamics and mean-field
picture\label{sec:dynamics-explanation}}

From the point of view of a network with couplings $\jj$ and and
biases $\hh$ whose evolution is described by a Glauber dynamics,
the bimodality and associated bistability appear because the couplings
$\jj$ are positive on average and symmetric, making the network an
excitatory one.

The positivity of the couplings appears because the average correlation
$\yccm$ between neurons is positive (\prettyref{fig:experimental_data}D).
But the symmetry of the couplings is also an essential factor. Consider
a neuron $i$ that on average projects negative couplings: $\sum_{j}^{j\ne i}\varLambda_{ji}<0$.
Such a neuron is ``inhibitory on average'' because its activation
will on average inhibit the neurons it is coupled to. But, owing to
coupling symmetry, ``inhibitory on average'' neurons are themselves
inhibited on average, not excited. Self-regulatory feedback loops,
possible in networks with asymmetric couplings, are impossible in
this case, and excitation can lead to regimes with a higher activity.
This phenomenon agrees with the known role of inhibitory neurons in
controlling low irregular activity in inhibition-dominated regimes
\citep{Vreeswijk96}.

\begin{figure}[H]
\includegraphics[width=1\textwidth]{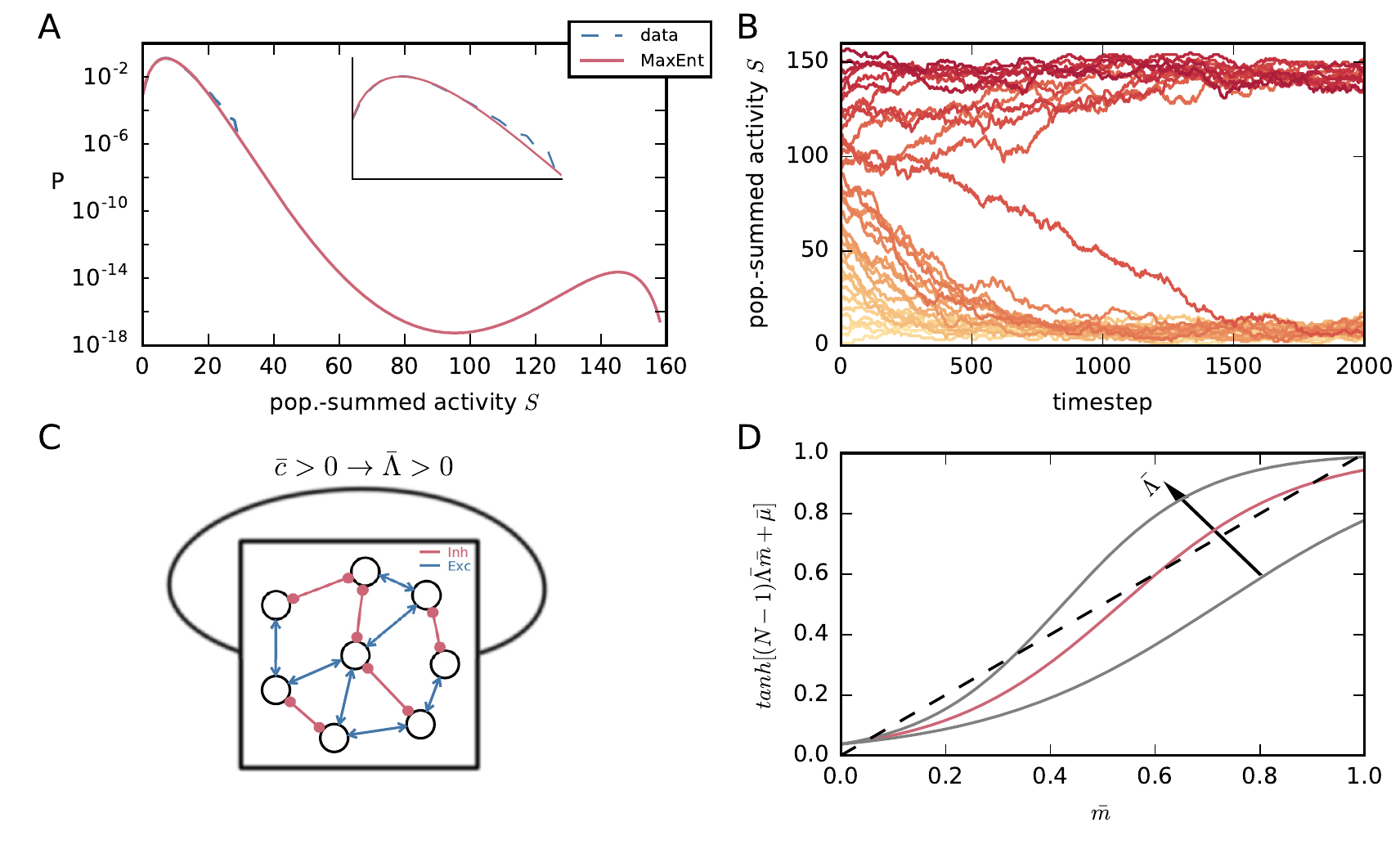}

\caption{\textbf{Reduced maximum-entropy model and mean-field picture.} (\textbf{A})
Red, solid: Probability distribution for the population-summed activity,
$\protect\PR(S)$ given by the reduced model for our dataset \prettyref{eq:pop-averaged-constraints_numerical_values};
note the two probability maxima. Blue, dashed: empirical distribution
of the population-summed activity from our dataset. (\textbf{B}) Population-summed
activities $S(t)$ obtained from several instances of Glauber dynamics
associated with the reduced model, with homogeneous couplings, $\varLambda_{ij}=\protect\yjjr$,
and biases, $\mu_{i}=\protect\yhhr$, of \prettyref{eq:Lagrange_values_reduced}.
As in \prettyref{fig:jump_longer_sampling}, each instance starts
with a different initial population activity $\protect\yss(0)$, having
different initial population sum $S(0)$, and is represented by a
different red shade, from $S(0)=0$ (light red) to $S(0)=N$ (dark
red). (\textbf{C}) Illustration of a self-coupled symmetric network
that is self-excitatory on average. Arrow-headed blue lines ($\to$)
represent excitatory couplings; circle-headed red lines ($\multimap$)
represent inhibitory couplings. (\textbf{D}) Self-consistency solution
of the naive mean-field equation, illustrated for different $\protect\yjjr$.
Larger $\protect\yjjr$ causes two additional intersections, corresponding
to one additional unstable and one additional stable solution. The
red curve corresponds to the $\protect\yjjr$ calculated from our
experimental data \prettyref{eq:Lagrange_values_reduced}.\label{fig:Appearance-of-bistable}}
\end{figure}

A naive mean-field analysis also confirms this. In the naive mean-field
approximation we imagine that each neuron is coupled to a field representing
the mean activities of all other neurons \citep{hartree1928}\citep[ch. 4]{negeleetal1988_r1998}\citep[ch. 6]{binneyetal1992_r2001}
(from the point of view of entropy maximization, we are replacing
the maximum-entropy distribution with one representing independent
activities, having minimal Kullback-Leibler divergence from the original
one \citep[chs 2, 16, 17]{opperetal2001b,amarietal2001,tanaka2001}).
Given the couplings $\jj$ and biases $\hh$, the mean activities
$\mm$ must satisfy $N$ self-consistency equations
\begin{equation}
\tanh(\sum_{j}^{j\ne i}\varLambda_{ij}m_{j}+\mu_{i})=m_{i}.\label{eq:self-consistency_vectorform}
\end{equation}
In the homogeneous case they reduce to the equation $\tanh[(N-1)\yjjr\ymmm+\yhhr]=\ymmm$
and correspond to the intersection of two functions of $\ymmm$: the
line $\ymmm\mapsto\ymmm$, and the curve $\ymmm\mapsto\tanh[(N-1)\yjjr\ymmm+\yhhr]$
that depends parametrically on $(\yhhr,\yjjr)$. See \prettyref{fig:Appearance-of-bistable}D:
for the Lagrange multipliers of our data, the curves these curves
intersect at two different values of $\ymmm$, meaning that there
are two solutions to the self-consistency equation, corresponding
to two different mean activities. These approximately correspond to
the maxima of the probability distribution for the population average
in \prettyref{fig:Appearance-of-bistable}A.

\subsubsection{Bistability ranges and population size}

Is the appearance of bimodality peculiar to our experimental dataset,
or can it be expected in other experimental datasets of neuronal activities?
Will it disappear at larger neuronal populations, or will it become
more prominent? We need to answer these questions to see whether this
is a general problem.

We again make an educated guess using reduced maximum-entropy model
and the distribution $\PR(\ysm\cond\yhhr,\yjjr)$, \prettyref{eq:probability-of-average}.
An elementary study of its convexity properties (second derivative)
shows that this distribution can have one minimum in the interior,
$0<\ysm<1$, or none, depending on the values of the parameters $(\yhhr,\yjjr)$.
The distribution has two probability maxima if it has one such minimum
for some value $\ysmin$, $0<\ysmin<1$. The conditions for this are
\begin{equation}
\frac{\di\PR(\ysm\cond\yhhr,\yjjr)}{\di\ysm}\biggr\rvert_{\mathrlap{\ysm=\ysmin}}=0,\quad\frac{\di^{2}\PR(\ysm\cond\yhhr,\yjjr)}{\di\ysm^{2}}\biggr\rvert_{\mathrlap{\ysm=\ysmin}}>0,\quad0<\ysmin<1,\label{eq:conditions_bistability}
\end{equation}
These conditions can be solved analytically and give the critical
ranges of multipliers $(\yhhr,\yjjr)$ for which bimodality occurs,
parametrically in $(\ysmin,\yjjr)$: 
\begin{equation}
\left\{ \begin{gathered}0<\ysmin<1,\\
\yjjr>\Psi'[1+(1-\ysmin)N]+\Psi'(1+\ysmin N),\\
\yhhr(\ysmin,\yjjr)=\yjjr/2-\ysmin N\yjjr-\Psi[1+(1-\ysmin)N]+\Psi(1+\ysmin N),
\end{gathered}
\right.\label{eq:bounds_bistability}
\end{equation}
where $\Psi(x)\coloneqq\di\ln\Gamma(x)/\di x$ and $\Gamma$ is the
Gamma function \citep[ch. 6]{abramowitzetal1964_r1972}\citep[chs 43, 44]{oldhametal1987_r2009}.
We then express the population-averaged single activity $\exper{\ysm}$
and Pearson correlation $\av{\rho}$, typically used in the literature,
in terms of $(\yhhr,\yjjr)$ using the definitions \prettyref{eq:correlation_pearson_pop-averages}
and the probability \prettyref{eq:probability-of-average}. Finally
we obtain the bimodality range for $\bigl(\exper{\ysm},\av{\rho}\bigr)$,
parametrically in $(\ysmin,\yjjr)$ within the bounds \prettyref{eq:bounds_bistability}.

\begin{figure}[H]
\includegraphics[width=1\textwidth]{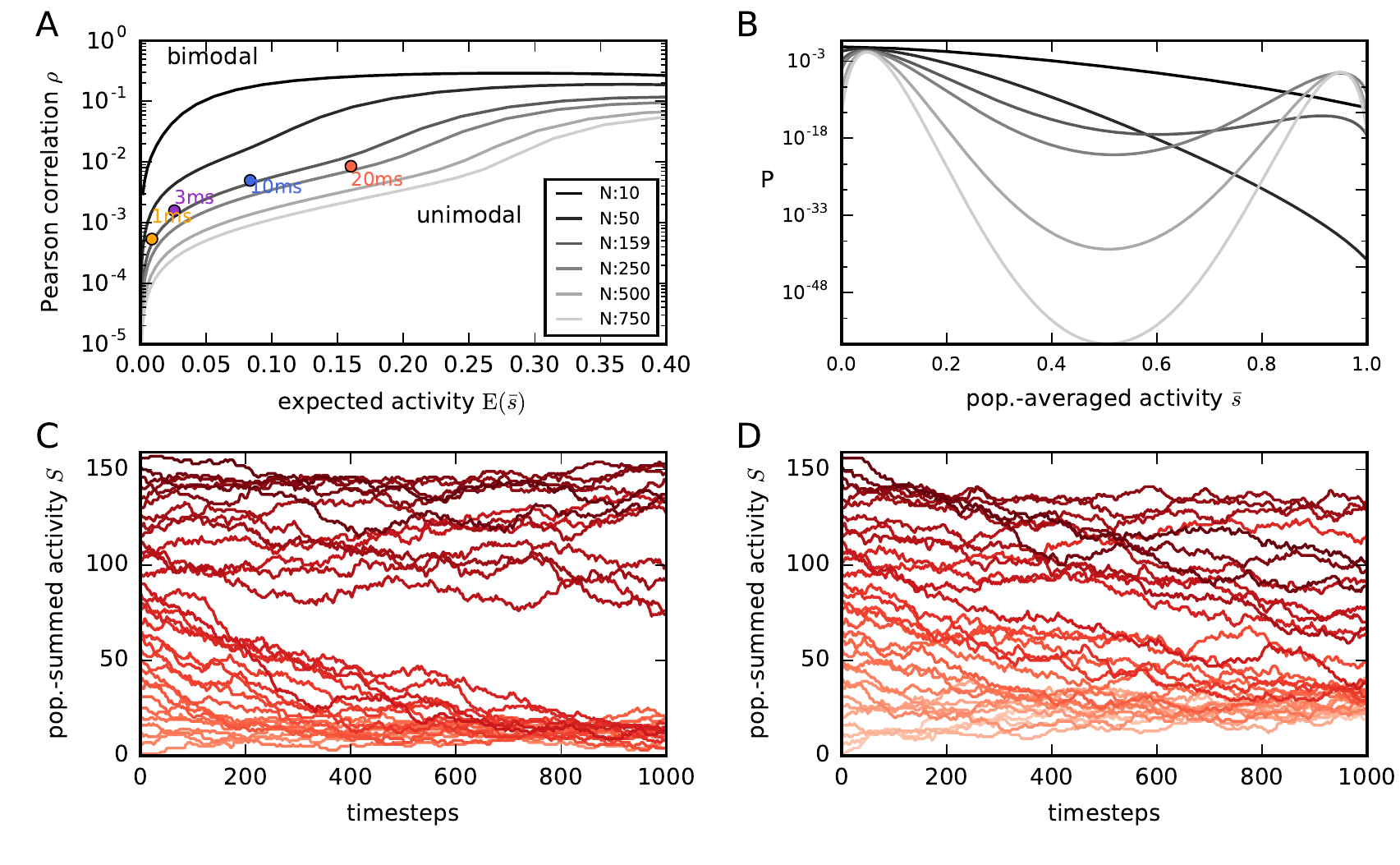}

\caption{\textbf{Bimodality ranges for the reduced model and effects of inhomogeneity.}
(\textbf{A}) The reduced maximum-entropy model \prettyref{eq:pairwise-reduced-maxent}
yields a distribution $\protect\PR(\protect\ysm)$ that is either
unimodal or bimodal, depending on the number of neurons $N$ and the
values of the experimental constraints $\bigl(\protect\exper{\protect\ysm},\protect\av{\rho}\bigr)$.
Each curve in the plot corresponds to a particular $N$ (see legend)
and separates the values $\bigl(\protect\exper{\protect\ysm},\protect\av{\rho}\bigr)$
yielding a unimodal distribution (below the curve) from those yielding
a bimodal one (above the curve). The curves are symmetric with respect
to $\protect\exper{\protect\ysm}=0.5$ (ranges $\protect\exper{\protect\ysm}>0.4$
not shown). Note how the range of constraints yielding bimodality
increases with $N$. Coloured dots show the experimental constraints
for our dataset, for different time-binnings with widths $\Delta=1\protect\ms$,
$\Delta=3\protect\ms$, $\Delta=10\protect\ms$, $\Delta=20\protect\ms$.
(\textbf{B}) Probability distributions of the reduced model for the
population-summed activity, $\protect\PR(S\protect\cond N)$, obtained
keeping the constraints \prettyref{eq:pop-averaged-constraints_numerical_values}
fixed and using different $N$ (same legend as panel A). (\textbf{C})
Population-summed activities $S(t)$ from several instances of Glauber
dynamics, all with the same normally-distributed couplings $\varLambda_{ij}$
and biases $\mu_{i}$, with means as in \prettyref{eq:Lagrange_values_reduced}
and \prettyref{fig:Appearance-of-bistable}B, and standard deviations
$\sigma(\varLambda_{ij})=0.009$, $\sigma(\mu_{i})=0.8$. Each instance
starts with a different initial population activity $\protect\yss(0)$,
having different initial population sum $S(0)$, and is represented
by a different red shade, from $S(0)=0$ (light red) to $S(0)=N$
(dark red). Note how the basins of attraction of the two metastable
regimes are wider than in the homogeneous case of \prettyref{fig:Appearance-of-bistable}B.
(\textbf{D}) The same as panel C, but with larger standard deviations\textbf{
}$\sigma(\varLambda_{ij})=0.019$, $\sigma(\mu_{i})=1.6$; the jumps
between the two metastable regimes become more frequent than in \prettyref{fig:Appearance-of-bistable}B,
indicating that the minimum between the modes becomes more shallow
with increasing inhomogeneity.\label{fig:m_rho}}
\end{figure}

The result is shown in \prettyref{fig:m_rho}A for various values
of $N$. A curve is associated with each $N$; values of $\bigl(\exper{\ysm},\av{\rho}\bigr)$
above such curves yield a bimodal distribution in the homogeneous
case.

Most important, \prettyref{fig:m_rho}A shows that the maximum-entropy
distribution will be bimodal for larger ranges of mean activities
and correlations, as the population size $N$ increases. Empirical
population-averaged quantities, on the other hand, should not change
with population size if they are sampled from a biologically homogeneous
neural population. This means that even if maximum-entropy does not
predict a bimodal distribution for the measured activities and correlations
of a particular small sample, it will predict a bimodal distribution
for a larger sample in a similar experimental setup. This phenomenon
is shown in \prettyref{fig:m_rho}B: keeping our constraints \prettyref{eq:pop-averaged-constraints_numerical_values}
fixed, when $N\lesssim150$ the distribution has only one maximum
for low activity, $\ysm\approx0.0497$, and when $N\gtrsim150$ a
second probability maximum for high activity, $\ysm\approx0.9502$,
appears. The probability at this second maximum increases sharply
until $N\approx200$ and thereafter maintains an approximately stable
value, roughly $6000$ times smaller than the low-activity maximum.
The minimum between the two modes becomes deeper and deeper as we
increase $N$ above $200$.

As mentioned in the previous section, exact studies with small samples
and studies with large samples and a different reduced model, which
takes into account the population-variance of the second moments,
indicate that the high-activity maximum in the inhomogeneous case
is larger (roughly $2000$ times smaller than the low-activity one
when $N=1000$) and shifted towards lower activities ($\ysm\approx0.25$
when $N=1000)$.

This can also be seen by by adding a Gaussian jitter to the multipliers
of the reduced case $\mu_{i}=\yhhr$, $\varLambda_{ij}=\yjjr$, making
it inhomogeneous. The results are shown in \prettyref{fig:m_rho}C--D.
The basin of attraction of the second metastable regime is shifted
to lower activities, and transitions between the two metastable regimes
become more likely for larger jitters. This means that inhomogeneity
makes the minimum in between the two modes shallower. The obtained
distribution is mathematically identical with the Boltzmann distribution
of the Sherrington \& Kirkpatrick infinite-range spin-glass \citep{sherringtonetal1975,kirkpatricketal1978}.
A more systematic analysis of the effect of inhomogeneity could therefore
employ methods developed for spin glasses \citep{fischeretal1991_r1993}.

The population-averaged activity and Pearson correlation of our data
(violet ``$3\ms$'' point in \prettyref{fig:m_rho}A) fall within
the bimodality range, as expected. The important question is whether
our dataset is a typical representative of this bimodality problem,
or an outlier. It is not an easy question to answer, as this kind
of experimental data are still rare, but we take as reference the
data summarized in Table~1 of Cohen \& Kohn \citep{cohenetal2011},
which reports firing rates and spike-count correlations $r_{\text{SC}}$.
The reported firing rates correspond to population-averaged activities
$\ymmm$ ranging between $0.02$ and $0.25$, if we use $3\ms$ time-bins.
We only need to estimate our Pearson correlation $\rho$ from their
spike-count correlation $r_{\text{SC}}$. Both are particular cases
of the ``cross-correlogram metric'' $r_{\text{CCG}}$ introduced
by Bair et al. \citep[App. A]{bairetal2001}: 
\begin{equation}
\begin{split}r_{\text{CCG}\,ij}(\tau) & \coloneqq\frac{\expe{\ynu_{i}(\tau)\,\ynu_{j}(\tau)}-\expe{\ynu_{i}(\tau)}\expe{\ynu_{j}(\tau)}}{\sqrt{[\expe{\ynu_{i}(\tau)^{2}}-\expe{\ynu_{i}(\tau)}^{2}]\,[\expe{\ynu_{j}(\tau)^{2}}-\expe{\ynu_{j}(\tau)}^{2}]}},\\
 & \text{with}\quad\ynu_{i}(\tau)\coloneqq\sum_{t=1}^{\tau}s_{i}(t),
\end{split}
\end{equation}
i.e. $\ynu_{i}(\tau)$ is the number of spikes of neuron $i$ during
the (real-)time window $\tau\varDelta$. This metric also equals the
area between times $-\tau\varDelta$ and $\tau\varDelta$ under the
cross-correlogram of neurons $i$ and $j$ (stationarity is assumed).
The spike count correlation $r_{\text{SC}}$ corresponds to $\tau=n\equiv T/\varDelta$,
and our Pearson correlation $\rho$ to $\tau=1$. Several studies
\citep{bairetal2001,mazureketal2002,kohnetal2005,smithetal2008,bakhurinetal2016}
report either measured values of $r_{\text{CCG}}(\tau)$ for different
windows $\tau$, or measured cross-correlograms. From their analysis
we can approximately say that $\rho\lesssim r_{\text{SC}}/20$, so
we take $\av{\rho}=\av{r_{\text{SC}}}/20$ as a safest-case value
(i.e. as far away from bimodality as possible).

\begin{figure}[h]
\begin{centering}
\includegraphics[width=0.9\textwidth]{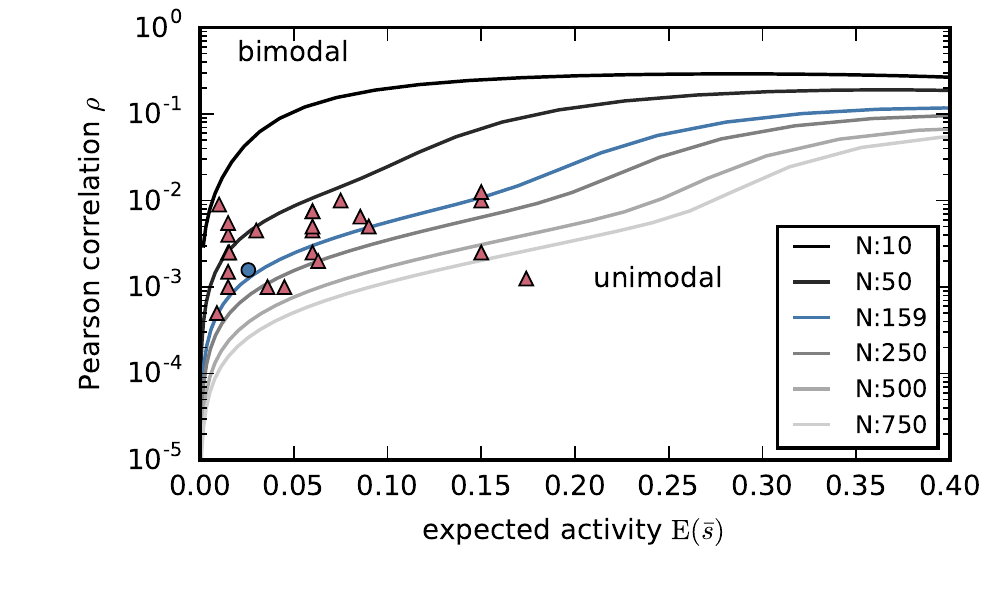}
\par\end{centering}

\caption{\textbf{Bimodality for experimental data from neuroscientific literature.}
Mean activities and correlations $\bigl(\protect\exper{\protect\ysm},\protect\av{\rho}\bigr)$
inferred from experimental data reported in Cohen \& Kohn \citep[Table 1]{cohenetal2011},
plotted upon the curves separating bimodal from unimodal maximum-entropy
distributions of \prettyref{fig:m_rho}A. The plot suggests that typical
experimental neural recordings of $250$ neurons and above are likely
to lead to bimodal maximum-entropy pairwise distributions\label{fig:cohen-points}.}
\end{figure}

Under these approximations the greatest part of the values summarized
by Cohen \& Kohn fall in the bimodality regions of \prettyref{fig:m_rho}A
if $N=250$, and almost all of them if $N=500$; see \prettyref{fig:cohen-points}.
These data points have only an indicative value but suggest that our
dataset is not an outlier for the bimodality problem. If those data
had been recorded from a population of $500$ neurons, they would
have yielded a bimodal pairwise maximum-entropy model. The bimodality
problem and its consequences need to be taken seriously. Is there
any way to eliminate it?

\subsection{Eliminating the bimodality: an inhibited maximum-entropy model and
Glauber dynamics}

We already mentioned in the Introduction why a maximum-entropy model
yielding a bimodal distribution in the population-summed activity
is problematic:
\begin{itemize}
\item The presence of two sharply distinct modes does not seem realistic
in view of present neuroscientific data, so the model is making unrealistic
predictions. The situation is even worse if the second mode peaks
at $90\%$ -- 90 neurons out of 100 simultaneously active!
\item As $N$ increases, the second mode becomes more pronounced, and the
minimum between the modes shallower: above a particular population
size, the bimodality cannot be dismissed as a small mathematical quirk.
\item The Boltzmann-learning procedure based on Glauber dynamics becomes
practically non-ergodic and the Lagrange multipliers of the model
are difficult or impossible to find.
\item The Glauber dynamics based on the pairwise model jumps between two
metastable regimes and cannot be used to generate realistic surrogate
data.
\item Finally, the fact that the position and height of the second mode
depend on $N$ (in the inhomogeneous case) goes against basic statistical
expectations. If we consider the $N$ neurons to be a sample, chosen
in an unsystematic way, of a larger population, then the maxima in
our probability assignments for the population averages of the sample
and of the larger population should roughly coincide (the former being
obtained from the latter by convolution with a hypergeometric distribution).
\end{itemize}
{}

We now propose a way to eliminate the bimodality and the above problems.
Let us re-examine what happens with the Glauber dynamics first.

\subsubsection{Importance of inhibition in neural networks: modified Glauber dynamics}

As mentioned in \nameref{sec:dynamics-explanation}, from the Glauber-dynamical
viewpoint jumps to high activities happen because the couplings $\jj$
are positive on average and symmetric, making the network an excitatory
one.

The positivity of the couplings is inevitable: it corresponds to an
experimentally observed positive average correlation. Their symmetry,
on the other hand, is a mathematical feature of the pairwise model
-- and, if we made an ungranted parallel with synaptic couplings,
it would not be a realistic feature.

Can we try to break this symmetry somehow? Can we add a minimal amount
of asymmetric inhibition to the Glauber dynamics?

\begin{figure}[H]
\includegraphics[width=1\textwidth]{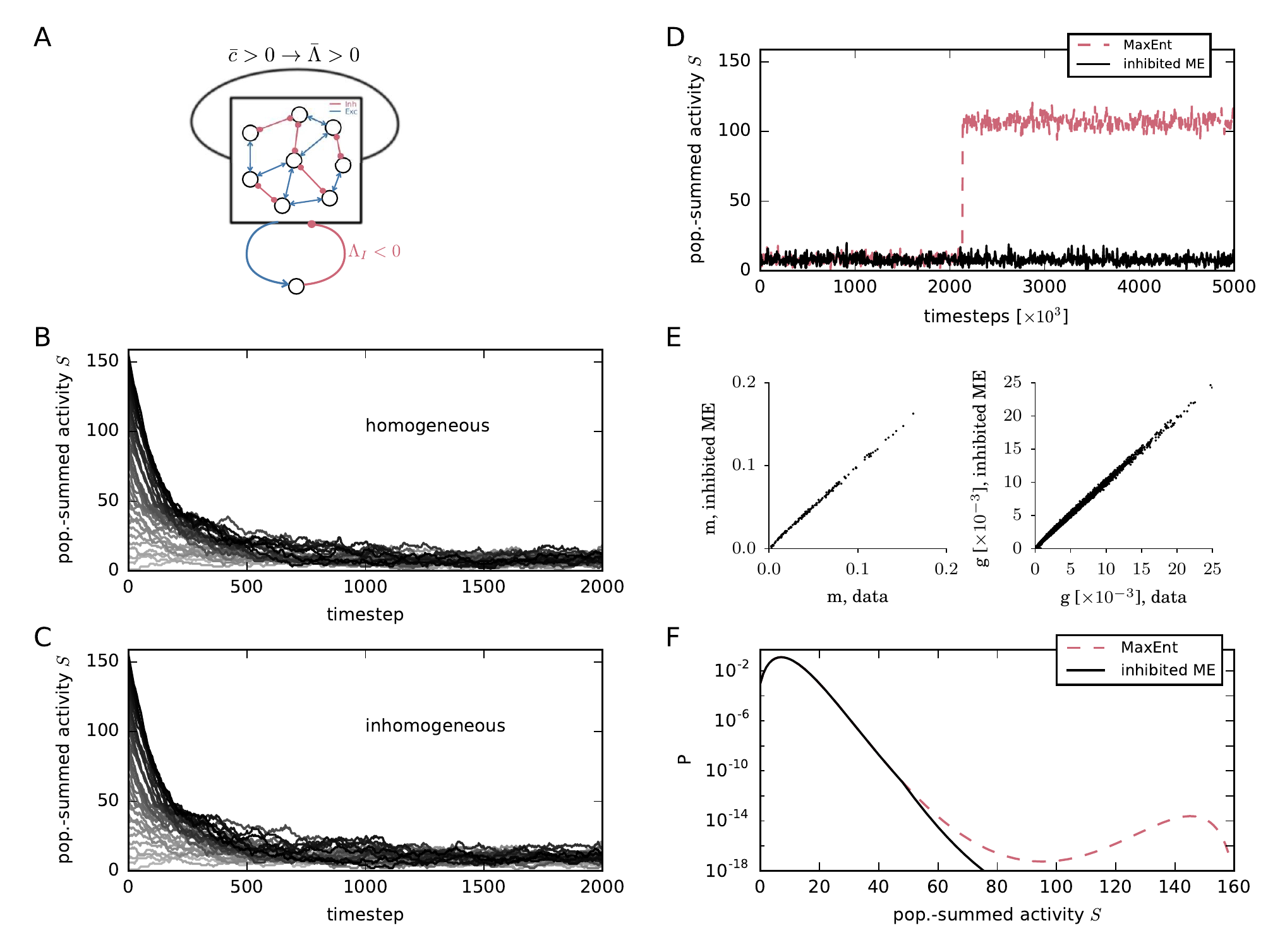}

\caption{\textbf{Asymmetric inhibition and elimination of bimodality and non-ergodicity.}
(\textbf{A}) Illustration of self-coupled network with additional
asymmetric inhibitory feedback. Each neuron receives inhibitory input
$\protect\yji<0$ from the additional neuron whenever the population-average
$\bar{s}$ becomes greater than the inhibition threshold $\theta$.
(\textbf{B}) Population-summed activities $S(t)$ from several instances
of the inhibited Glauber dynamics, with $\protect\yji=-25$, $\theta=0.3$,
and same homogeneous $\varLambda_{ij}=\protect\yjjr$, $\mu_{i}=\protect\yhhr$
of \prettyref{eq:Lagrange_values_reduced}, as used for \prettyref{fig:Appearance-of-bistable}B.
Each instance starts with a different initial population activity
$\protect\yss(0)$, having different initial population sum $S(0)$,
and is represented by a different grey shade, from $S(0)=0$ (light
grey) to $S(0)=N$ (black). Note the disappearance, thanks to inhibition,
of the bistability that was evident in the ``unhinibited'' case
of \prettyref{fig:Appearance-of-bistable}B. (\textbf{C}) Analogous
to panel B, with $\protect\yji=-25$, $\theta=0.3$, but inhomogeneous
normally distributed couplings and biases as in the unhinibited case
of \prettyref{fig:m_rho}C. Note again the disappearance, thanks to
inhibition, of the bistability that was evident in the activities
$S(t)$ of that figure. (\textbf{D}) Comparison of a longer ($5\times10^{6}$
timesteps) Glauber sampling with couplings and biases of \prettyref{fig:first_Boltzmann}D
obtained from our first Boltzmann learning, and inhibited-Glauber
sampling with same couplings and biases and $\protect\yji=-25$, $\theta=0.3$.
The comparison confirms that inhibition eliminates the second metastable
regime and makes the Glauber dynamics ergodic. (\textbf{E}) Time averages
$m_{i}$ and $g_{ij}$ obtained from Boltzmann learning for the inhibited
model $\protect\PI$, versus experimental ones. (\textbf{F}) Probability
distribution of the population-summed activity $\protect\PI(S)$ given
by the inhibited model \prettyref{eq:inhibited_pairwise_maxent} for
our dataset \prettyref{eq:pop-averaged-constraints_numerical_values},
compared with the one previously given by the reduced model $\protect\PR(S)$,
\prettyref{fig:Appearance-of-bistable}A. Asymmetric inhibition, expressed
by the reference prior \prettyref{eq:reference_prior_minimum-relative-entropy},
has eliminated the second mode.\label{fig:Inhibitory-neurons-to}}
\end{figure}

The answer is yes, in a very simple way: by connecting all $N$ neurons
to a single inhibitory neuron that instantaneously activates whenever
their average activity exceeds a threshold $\theta$, having a value
in the set $\set{1/N,2/N,\dotsc,(N-1)/N}$ (the cases $\theta=0\text{ or }1$
are trivial). Upon activation, the inhibitory neuron sends inhibitory
feedback, $\yji<0$, to all other $N$ neurons (see \prettyref{fig:Inhibitory-neurons-to}A).
The algorithm for this ``inhibited'' Glauber dynamics (including
how the ``instantaneously'' is implemented) is explained in the
Materials and Methods section.

The results from simulations of the inhibited Glauber dynamics are
shown in \prettyref{fig:Inhibitory-neurons-to}C--D; in all cases
the inhibitory coupling $\yji=-24.7$ and the inhibition threshold
$\theta=0.3$. In the reduced homogeneous case the couplings $\varLambda_{ij}=\yjjr$
and biases $\mu_{i}=\yhhr$, \prettyref{eq:Lagrange_values_reduced},
are the same that led to bistability in \prettyref{fig:Appearance-of-bistable}B;
in the inhomogeneous case they are the same, normally distributed,
that led to bistability in \prettyref{fig:m_rho}C--D. In either case,
the additional inhibitory neuron has eliminated the bistability, leaving
only the stable low-activity regime.

Furthermore, also in the case of the inhomogeneous couplings and biases
(distributed as in \prettyref{fig:first_Boltzmann}D) that caused
the quasi-non-ergodic behaviour in our first Boltzmann learning results,
\prettyref{fig:jump_longer_sampling}, the addition of the inhibitory
neuron (again with $\yji=-24.7$, $\theta=0.3$) eliminates the second
metastable state: see \prettyref{fig:Inhibitory-neurons-to}E.

In summary: the asymmetric coupling of an additional inhibitory neuron
clearly eliminates the bistability of the Glauber dynamics. This works
for any network size $N$ with an appropriate choice of the inhibitory
coupling $\yji<0$ and threshold $\theta$.

We now show that this idea also eliminates our original problem: the
bimodality of the pairwise maximum-entropy model.

\subsubsection{Inhibited maximum-entropy model\label{sub:Modified-maximum-entropy}}

The pairwise maximum-entropy model is the stationary distribution
of the Glauber dynamics with symmetric couplings. We have now modified
the latter in an asymmetric way. The stationary distribution of the
inhibited Glauber dynamics of \prettyref{fig:Inhibitory-neurons-to}
cannot, therefore, be a pairwise maximum-entropy model. It turns out,
however, that \emph{it is still a maximum-entropy model}, of the following
form: 
\begin{equation}
\begin{split}\PI(\yss\cond\hh,\jj,\yji,\theta) & =\begin{aligned}[t] & \frac{1}{\zi(\hh,\jj,\yji,\theta)}\times\\
 & \exp\bigl[\sum_{i}\mu_{i}s_{i}+\sum_{i>j}\varLambda_{ij}s_{i}s_{j}+\yji N\,\yG(\ysm-\theta)\bigl],
\end{aligned}
\\
\zi(\hh,\jj,\yji,\theta) & \coloneqq\sum_{\yss}\exp\bigl[\sum_{i}\mu_{i}s_{i}+\sum_{i>j}\varLambda_{ij}s_{i}s_{j}+\yji N\,\yG(\ysm-\theta)\bigl],\\
G(\ysm-\theta) & \coloneqq(\ysm-\theta)\,\HS(\ysm-\theta),
\end{split}
\label{eq:inhibited_pairwise_maxent}
\end{equation}
where $\yji$ is the (negative, in our case) coupling strength from
the inhibitory neuron to the other neurons, $\theta$ is the activation
threshold of the inhibitory neuron, and $\HS$ is the Heaviside step
function. We call \prettyref{eq:inhibited_pairwise_maxent} the \emph{inhibited
pairwise maximum-entropy model.}

\begin{figure}[h]
\includegraphics[width=0.9\textwidth]{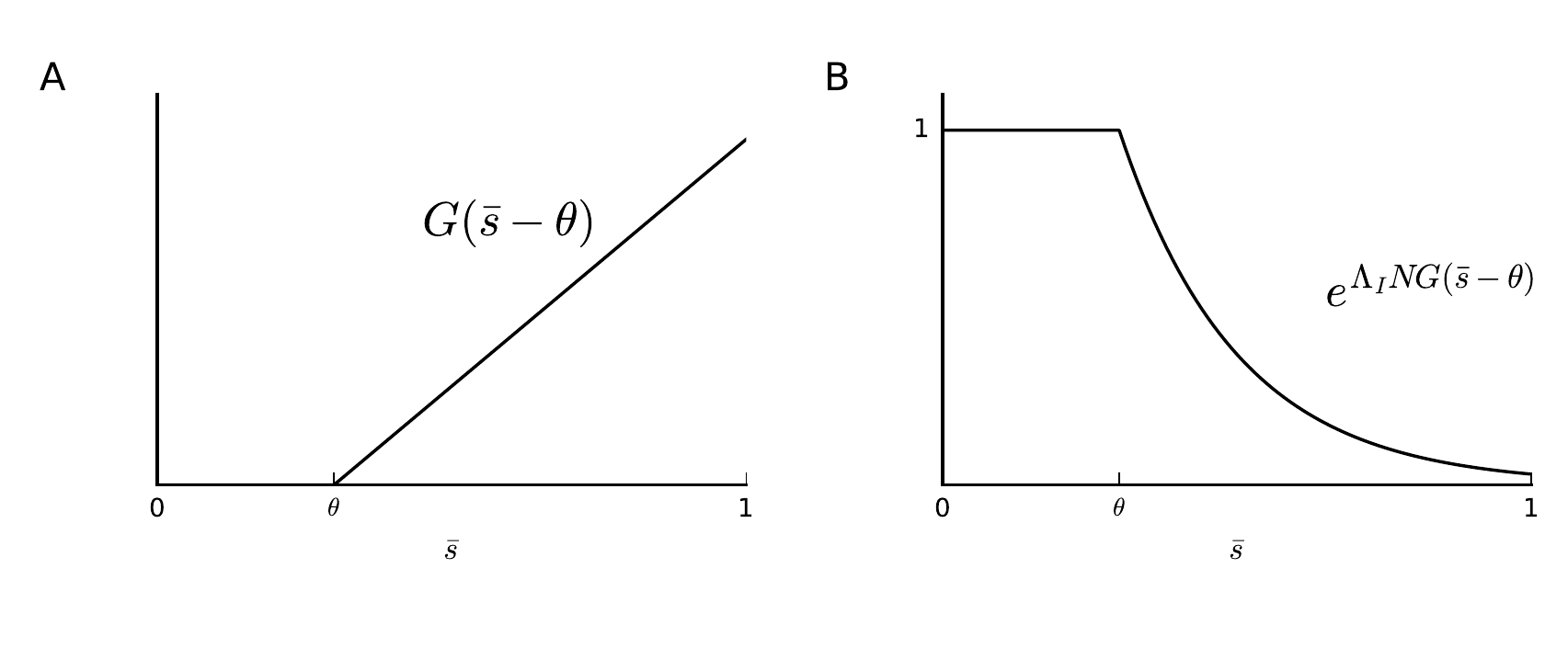}

\caption{\textbf{Reference prior. }The function $\protect\yG(\protect\ysm-\theta)$
and its exponential\label{fig:function_G}}
\end{figure}

The function $\yG(\ysm-\theta)$ (plotted in \prettyref{fig:function_G}
together with its exponential) can also be written as a linear combination
of population-averaged $K$-tuple activities, $s_{i_{1}}s_{i_{2}}\dotsm s_{i_{K}}$,
for $K$ equal to $N\theta$ and larger (we leave the proof of this
as a classic ``exercise for the reader''):
\begin{equation}
N\,\yG(\ysm-\theta)=\sum_{K=N\theta}^{N}\binom{-N\theta}{-K+1}\,\left(\sum_{i_{1}<i_{2}<\dotsb<i_{K}}s_{i_{1}}s_{i_{2}}\dotsm s_{i_{K}}\right),\label{eq:G_as_products}
\end{equation}
the linear coefficients being binomial coefficient functions \citep{fowler1996},
which have alternating signs. For example, if $N=5$ and $\theta=3/5$,
\begin{multline}
N\,\yG(\ysm-\theta)={}\\
(s_{2}s_{3}s_{4}s_{5}+s_{1}s_{3}s_{4}s_{5}+s_{1}s_{2}s_{4}s_{5}+s_{1}s_{2}s_{3}s_{5}+s_{1}s_{2}s_{3}s_{4})-{}\\
3\,s_{1}s_{2}s_{3}s_{4}s_{5}.
\end{multline}
(This function differs from the additional function appearing in maximum-entropy
model by Tkačik et al. \citep{tkaciketal2013,tkaciketal2014,tkaciketal2014b},
which consists in $N+1$ constraints enforcing the observed population-average
distribution. For reasons discussed at the end of \prettyref{sub:Pairwise-maximum-entropy-model},
the use of all those constraints may not be justified or meaningful.)

The stationarity of the distribution $\PI(\yss)$ under the inhibited
Glauber dynamics is proved in the Materials and Methods section. This
distribution is a maximum-entropy model in two different ways:
\begin{enumerate}[label=(\alph*)]
\item As an application of the minimum-relative-entropy (minimum-discrimination-information)
principle \citep{kullback1959_r1978,kullbacketal1984,csiszar1985,campbell1985,kullbacketal1951,jaynes1963,renyi1961,chencov1972_t1982,hobsonetal1973,aczeletal1975,gulletal1984,gulletal1984b,kullbacketal1987,kullback1987,gulletal1990_r1999},
with the pairwise constraints \prettyref{eq:constraints_single_and_couple},
with respect to the reference (or prior) probability distribution
\begin{equation}
P_{0}(\yss\cond\yji,\theta)\propto\exp\bigl[\yji N\,\yG(\ysm-\theta)\bigr],\qquad(N\theta\in\set{1,2,\dotsc,N-1}).\label{eq:reference_prior_minimum-relative-entropy}
\end{equation}
This distribution assigns decreasing probabilities to states with
average activities above $\theta$; see \prettyref{fig:function_G}B.
This probability can be interpreted as arising from a more detailed
model in which we know that external inhibitory units make activities
above the threshold $\theta$ increasingly improbable (we explain
this in the Discussion section). In this interpretation the parameters
$\yji$ and $\theta$ are chosen a priori.
\item As an application of the ``bare'' maximum-entropy principle, given
the pairwise constraints \prettyref{eq:constraints_single_and_couple}
and an additional constraint for the expectation of $N\,\yG(\ysm-\theta)$:
\begin{multline}
\expebi{N\,\yG(\ysm-\theta)}=\sum_{S=N\theta}^{N}(S-N\theta)\,\PI(S)\\
{}=\sum_{K=N\theta}^{N}\binom{-N\theta}{-K+1}\,\expebi{\sum_{i_{1}<i_{2}<\dotsb<i_{K}}s_{i_{1}}s_{i_{2}}\dotsm s_{i_{K}}}.\label{eq:new_constraint_function}
\end{multline}
This is a constraint of the ``tail first moment'', so to speak,
of the probability for the population-averaged activity $\PI(\ysm)$:
it determines whether the right tail of $\PI(\ysm)$ has a small ($\yji<0$)
or heavy ($\yji>0$) probability. It can also be seen as a constraint
on the $N\theta$th and higher moments, owing to \prettyref{eq:G_as_products}.
In this interpretation the parameter $\yji$ is the Lagrange multiplier
associated with this constraint, hence it is determined by the data;
the parameter $\theta$ is still chosen a priori. Note, however, that
experimental data are likely to give a vanishing time average of $N\,\yG(\ysm-\theta)$,
so that $\yji=-\infty$. This interpretation has therefore to be used
with care, for the reasons discussed at the end of \prettyref{sub:Pairwise-maximum-entropy-model}.
\end{enumerate}
The inhibited model $\PI$ includes Shimazaki's model \citep{shimazakietal2015}
and its ``simultaneous silence'' constraint as the limit $\yji\to-\infty$,
$\theta=1/N$. Because of this limit, Shimazaki's model has a sharp
jump in probability when $\ysm=1/N$ (the constraint uniformly removes
probability from $P(\ysm>1/N)$ and gives it to $P(\ysm=0)$), whereas
the inhibited model $\PI$ only presents a kink when $\ysm=\theta$,
with a discontinuity in the derivative proportional to $\yji$.

Several features of the inhibited maximum-entropy model \prettyref{eq:inhibited_pairwise_maxent}
are worth remarking upon:
\begin{enumerate}
\item The inhibited distribution $\PI$ includes the pairwise one $\PP$,
\prettyref{eq:pairwise-maxent}, as the particular case $\yji=0$
(obviously, as this is equivalent to removing the inhibitory neuron).
\item \label{enu:behaviour_inhibited_below_threshold}Pairwise and inhibited
distributions $\PP$ and $\PI$ having same Lagrange multipliers $(\hh,\jj)$
are\emph{ equal }if restricted to states with population-averaged
activity below the threshold $\theta$, because $\yG(\ysm-\theta)=0$
if $\ysm\leqslant\theta$: 
\begin{equation}
\PP(\yss\cond\hh,\jj,\ysm\leqslant\theta)=\PI(\yss\cond\hh,\jj,\yji,\theta,\ysm\leqslant\theta).\label{eq:equality_truncated_distributions}
\end{equation}
Said otherwise, the pairwise and inhibited distributions have the
same shape for $\ysm\leqslant\theta$, modulo rescaling by a constant
factor:
\begin{equation}
\PI(\yss\cond\hh,\jj,\yji,\theta)=\PP(\yss\cond\hh,\jj)\times\frac{\zp(\hh,\jj)}{\zi(\hh,\jj,\yji,\theta)},\quad\ysm\leqslant\theta.\label{eq:same_shape_pairwise_inhibited}
\end{equation}

\item \label{enu:behaviour_inhibited_above_threshold}For states with average
activity $\ysm$ above the threshold $\theta$, the inhibited model
is a ``squashed'' version of the pairwise one when $\yji<0$:
\begin{equation}
\PI(\yss\cond\hh,\jj,\yji,\theta)\propto\PP(\yss\cond\hh,\jj)\times\exp\bigl[N\yji\,(\ysm-\theta)\bigr],\quad\ysm>\theta.
\end{equation}

\item If $\yji\ne0$, then inhibited and pairwise models with the same Lagrange
multipliers $(\hh,\jj)$ have \emph{different} expectations for single
and coupled activities:
\begin{equation}
\begin{aligned}\expei{s_{i}\cond\hh,\jj,\yji,\theta} & \ne\expep{s_{i}\cond\hh,\jj},\\
\expei{s_{i}s_{j}\cond\hh,\jj,\yji,\theta} & \ne\expep{s_{i}s_{j}\cond\hh,\jj},
\end{aligned}
\end{equation}
and obviously also different covariances and correlations.
\end{enumerate}
Remarks \ref{enu:behaviour_inhibited_below_threshold} and \ref{enu:behaviour_inhibited_above_threshold}
above imply that if the inhibitory coupling $\yji$ is negative and
very large, so that $\exp\bigl[N\yji(\ysm-\theta)\bigr]\approx0$
when $\ysm>\theta$, then \emph{the inhibited maximum-entropy distribution}
$\PI$ \emph{is approximately equal to the truncated distribution}
$\PT$ -- the incorrect one \prettyref{eq:effective_truncated_P}
obtained via Boltzmann learning -- having the same multipliers $(\hh,\jj)$
and threshold $\theta$:
\begin{equation}
\PI(\yss\cond\hh,\jj,\yji,\theta)\approx\PT(\yss\cond\hh,\jj,\theta)\quad\text{if \ensuremath{\yji\ll-1}},
\end{equation}
(mathematically speaking we have pointwise convergence as $\yji\to-\infty$),
and their expectations are also approximately equal. This suggests
a way to reinterpret and keep the results of our first Boltzmann-learning
algorithm \prettyref{fig:first_Boltzmann}.

\subsubsection{Boltzmann learning for the inhibited maximum-entropy model}

Our first Boltzmann-learning calculation, with results shown in \prettyref{fig:first_Boltzmann},
returned a distribution that reproduced the desired constraints $(\mm,\ygg)$.
But that distribution turned out to be not the true pairwise maximum-entropy
one, but a truncated version of it $\PT$, \prettyref{eq:effective_truncated_P},
owing to the bimodality of the true pairwise distribution and the
resulting non-ergodicity.

If we had we decided to apply the inhibited pairwise maximum-entropy
model $\PI(\yss\cond\hh,\jj,\yji,\theta)$ (with $\yji\ll-1$ and
$0.3\lesssim\theta\lesssim0.5$) to our data, instead of the pairwise
one $\PP(\yss\cond\hh,\jj)$, and had sought its Lagrange multipliers
via Boltzmann learning, then the results would have been the same
as in \prettyref{fig:first_Boltzmann}. It is clear why: the inhibitory
neuron would not have allowed jumps to higher activities, unlike \prettyref{fig:jump_longer_sampling}A
(see \prettyref{fig:Inhibitory-neurons-to}D); and at the same time
it would not have influenced the dynamics below $\ysm\approx0.3$
(i.e. $S\approx50$). The sampling phase of our Boltzmann learning
would have been sufficient. We confirm this by applying Boltzmann-learning
procedure (with the inhibited Glauber dynamics) to find the multipliers
of the inhibited model with $\yji=-25$, $\theta=0.3$. The resulting
multipliers are close to those in \prettyref{fig:first_Boltzmann}D,
and the constraints are satisfied, see \prettyref{fig:Inhibitory-neurons-to}E.

Results obtained with a non-ergodic Boltzmann learning, and therefore
incorrect for the pairwise maximum-entropy model, can therefore be
reinterpreted as correct results for the inhibited pairwise maximum
entropy model, with appropriately chosen $\yji$ and $\theta$. This
is important for any work in the literature that may unknowingly have
been affected by non-ergodicity.

\section{Discussion}

\subsection{Summary}

The pairwise maximum-entropy model, applied to experimental neuronal
data of populations of $200$ and more neurons, is very likely to
give a \emph{bimodal} probability distribution for the population-averaged
activity, $P(\yss)$. We have provided evidence for this claim in
\prettyref{sub:The-problem-bimodality}, starting from an experimental
dataset and then looking at summarized data from the literature. The
first mode is the one observed in the data. The second mode (unobserved)
can appear at very high activities (even $90\%$ of the population
simultaneously active) and its height increases with population size.

The presence of a second mode is problematic for several reasons:
\begin{itemize}
\item As far as we know, a second mode has never been observed in experimental
recordings, and surely not at high activity -- data in which 180 out
of 200 neurons spike simultaneously are unheard of. So it is an unrealistic
prediction of the pairwise model.
\item Above certain population sizes the second mode cannot be dismissed
as too small to be recorded, because it becomes more pronounced as
$N$ increases, and the minimum that separates it from the main mode
becomes shallower.
\item The Boltzmann-learning \citep{ackleyetal1985,hintonetal1986_r1999,brodericketal2007}
procedure based on asynchronous Glauber dynamics \citep[chap. 29]{glauber1963d,mackay1995_r2003}
becomes practically non-ergodic -- it can already be so for population
sizes of roughly 50 neurons -- so that the Lagrange multipliers of
the pairwise model are difficult or impossible to find. Approximate
methods like mean-field \citep{hartree1928,negeleetal1988_r1998,opperetal2001},
Thouless-Anderson-Palmer \citep{thoulessetal1977,opperetal2001},
Sessak-Monasson \citep{sessaketal2009,sessak2010}   also seem to
break down in this case.
\item The Glauber dynamics based on the pairwise model jumps between two
metastable regimes, remaining in each for long times (owing to its
asynchronous update) and cannot be used to generate realistic surrogate
data.
\item The fact that the position and height of the second mode vary with
$N$ contradicts the natural assumption that the recorded $N$ neurons
are a ``random sample'' of a larger population. (The probability
calculus tells us that the population-average distributions of a full
population and a ``random sample'' from it should have maxima at
roughly the same relative heights and locations, since they are connected
by convolution with a hypergeometric distribution \citep[ch. II]{feller1950_r1968}\citep[ch. 4]{ross1976_r2010}\citep[ch. 3]{jaynes1994_r2003}\citep[cf. also][]{wolpertetal1994,wolpertetal1994b,wolpertetal1995,ghoshetal1997,portamanaetal2015}.)
\end{itemize}
Eliminating the second mode also eliminates all these problems.

We gave an intuitive explanation of why the second mode appears: because
the pairwise model, given positive pairwise correlations, corresponds
to a network that is excitatory on average and symmetric. And symmetric
connectivity is incompatible with the presence of a subset of neurons
that have an inhibitory effect, but receive excitatory input (see
\prettyref{sec:dynamics-explanation}). This explanation also suggested
a way to eliminate the second mode: by adding a minimal asymmetric
inhibition to the network, in the guise of an additional, asymmetrically
coupled inhibitory neuron (\prettyref{fig:Inhibitory-neurons-to}A).

This idea led to the construction of an ``inhibited'' pairwise maximum-entropy
model $\PI(\yss)$, \prettyref{eq:inhibited_pairwise_maxent}, with
the important properties:
\begin{itemize}
\item It is a maximum-entropy or minimum-relative-entropy model.
\item It is the stationary distribution of a particular asynchronous Glauber
dynamics with pairwise couplings.
\item Its Lagrange multipliers can be found via Boltzmann learning.
\item Its parameters can be chosen to have the main mode only.
\item It is numerically equal to the distribution one would obtain from
a non-ergodic Boltzmann learning.
\end{itemize}

\subsection{In defence of the inhibited model}

We have already argued at length that bimodality is a problem in the
application of the pairwise model, and do not dwell on this in this
discussion. We wish to stress, though, that the presence of bimodality
and non-ergodicity can easily go unnoticed. We urge researchers who
use Boltzmann learning or one of the mentioned approximations to check
for the presence of bimodality and non-ergodicity by starting the
sampling from different initial conditions, at low and high activities,
looking out for bistable regimes \citep[cf.][\S\ 2.1.3]{landauetal2000_r2015}.
One way out of this problem is to use other sampling techniques or
Markov chains different from the Glauber one \citep{mackay1995_r2003,landauetal2000_r2015,binder1997,binder1984_r1987}.

Different readers will draw different conclusions from the presence
of bimodality. Some may dismiss or abandon the whole pairwise model
as flawed. Some may still want to use it, bimodality notwithstanding.
Some may look for other maximum-entropy-inspired alternatives. We
have presented \emph{one} (as opposed to \emph{the}) such alternative:
the ``inhibited'' pairwise maximum-entropy model $\PI$, \prettyref{eq:inhibited_pairwise_maxent}.
It is an interesting alternative for at least two reasons.

First, the inhibited distribution $\PI$ is stationary under a Glauber
dynamics with \emph{pairwise} couplings. Consider that pairwise models
with additional constraints are stationary under Gibbs samplers with
higher-order couplings -- and thus lose some of their analogies with
real neuronal networks.

Second, the inhibited distribution $\PI$ incorporates the effects
of neural inhibition in a simple way. These effects are represented
by the reference or prior probability $P_{0}(\yss\cond\yji,\theta)$,
\prettyref{eq:reference_prior_minimum-relative-entropy}.

Some readers may actually object to the usefulness of the inhibited
distribution $\PI$ exactly because it is derived from a particular
prior via minimum-relative-entropy, and may thus appear less ``non-committal''
or less ``unstructured'' than a ``bare'' maximum-entropy one.
We would like to briefly counter this argument by pointing out that
bare maximum-entropy can be quite ``committal'', and that reference
priors can correct that.

The statement ``the maximum-entropy method gives the maximally non-commital
probability distribution consistent with the given information''
and variations thereof are frequently repeated in the literature.
But there are many qualifications behind this statement, especially
behind the terms ``non-committal'' and ``information''. The term
``information'' does not mean only ``experimental data'': it also
means knowledge of the \emph{assumptions underlying }the specified
problem and the variables implied. The way we set up a maximum-entropy
problem implies many underlying assumptions, already before experimental
data are taken into account \citep{jaynes1986d_r1996,jaynes1986b}.

A concrete assumption underlying the bare maximum-entropy principle
applied to neuronal activity is that \emph{the recorded neurons are
not a sample from a larger population of neurobiologically similar
neurons}. It is easy to expose this assumption in the homogeneous
case. If we assume that our $N$ neurons are a sample from a larger
population, the maximum-entropy principle requires that the moment
constraints be applied to the average of the \emph{full} population,
not of the sample \citep[§ 3.2]{portamanaetal2015}. The marginal
distribution of the sample will \emph{not} be a maximum-entropy distribution.
The assumption above is quite strong and neurobiologically unrealistic,
but does not seem to have bothered researchers who applied maximum-entropy
to samples; or maybe it escaped their attention. In any case it shows
that the bare maximum-entropy principle is far from ``non-committal''
or ``unstructured''.

The ``committal'' nature of bare maximum-entropy also appears in
its derivation from the probability calculus. This derivation requires
a particular prior \citep{csiszar1984,csiszar1985,jaynes1986d_r1996,jaynes1986b,portamana2009},
but one could use other, quite natural priors (e.g., the ``flat prior
over probability distributions'' considered by Bayes \citep[Scholium]{bayes1763}
and Laplace \citep[p. xvii]{laplace1814_r1819}) and the result would
\emph{not }be a bare maximum-entropy distribution.

Reference priors can in some cases correct such implicit assumptions.
For example, consider a pair of neurons with binary states $s_{1}$
and $s_{2}$. Without ``experimental data'', the maximum-entropy
principle assigns a uniform probability of $1/4$ to each of the four
possible joint states $(s_{1},s_{2})$. The probability assigned to
the total activity, $S\coloneqq s_{1}+s_{2}$, is therefore \emph{not}
\emph{uniform} ($2/4$ probability to $S=1$ and $1/4$ probability
to the remaining two values). If we apply the maximum-entropy principle
to the total activity $S$ directly, instead, it gives a \emph{uniform}
probability of $1/3$. Both applications of the principle are consistent,
but they use different assumptions about the structure of the biophysical
problem. The information implicit in the first application can be
specified in the second by using the minimum-relative-entropy method
with a non-uniform prior distribution assigning $2/4$ probability
to $S=1$. (Something analogous happens in statistical mechanics with
the probability distribution for energy, in which a ``density of
states'' term multiplies the Boltzmann factor). Some implicit assumptions,
however, like the sampling assumption previously discussed, cannot
be corrected by reference priors.

The necessity of reference priors, reflecting deeper assumptions,
is well-known in maximum-entropy image reconstruction \citep{skilling1986,weir1991},
for example of astronomical sources \citep{skillingetal1984,weir1991}:
as Skilling remarked, ``bare maximum-entropy is surprised to find
isolated stars, but astronomers are not'' \citep{jaynes1986d_r1996}.

An analogous remark can be made in our case: bare maximum-entropy
is surprised to find so many inactive neurons, and it tries to make
some more active ones by creating a second maximum, if that does not
break the constraints. But neuroscientists are not surprised at inactive
neurons. Bare maximum-entropy assumes that we have abstract ``units''
whose states are symmetrically exchangeable. But neuroscientists know
that these units are \emph{neurons,} whose individual and collective
properties are \emph{asymmetric} with respect to state exchanges,
for biophysical reasons. The prior of the inhibited model $\PI$ reflects
this asymmetry. It is fortunate that we can partially correct the
symmetry assumption of bare maximum-entropy by using a prior, without
having to overturn our whole space of variables.

The long argument above shows, we hope, that the inhibited model $\PI$
and its reference prior do not break the ``non-committal'' nature
of the maximum-entropy principle; rather, they prevent maximum-entropy
from committing to unrealistic assumptions. The inhibited model can
therefore be quite useful as a realistic hypothesis against which
to check or measure the prominence of correlations in simulated or
recorded neural activities.

\subsection{Back to the big picture}

Let us conclude by returning to the general modelling point of view
outlined in the Introduction. Our original goal was to simplify $P(\text{activity}\cond\text{stimuli})$
via a set of intermediate models $\set{M}$, each with a multi-dimensional
parameter $\bm{\alpha}$, by $P(\text{activity}\cond\text{state})=\sum_{M}\int P(\text{activity}\cond\bm{\alpha},M)\,P(\bm{\alpha},M\cond\text{state})\,\di\bm{\alpha}$.
They should be neurobiologically sound but mathematically manageable.
The next step is to establish which of these models is most probable,
given the observed activities:
\begin{equation}
P(\bm{\alpha},M\cond\text{activity})\propto P(\text{activity}\cond\bm{\alpha},M)\,P(\bm{\alpha},M),
\end{equation}
where $P(\bm{\alpha},M)$ are the prior probabilities we assign to
these models. The determining factor is usually $P(\text{activity}\cond\bm{\alpha},M)$,
called the \emph{evidence}. This last step is adamantly explained
and discussed in a beautiful paper by Mackay \citep{mackay1992}.
The pairwise maximum-entropy model $\PP$, and the inhibited pairwise
maximum-entropy model $\PI$ presented in this paper, are two examples
of such ``$M$''. And we can consider higher-moment maximum-entropy
models, models with deeper underlying assumptions, and even models
not based on maximum-entropy at all. It is against this last step
that the question of the importance of pairwise and higher-order correlations,
and of maximum-entropy models in general, acquires its full meaning
and can be given a precise answer.

\section{Materials and Methods}

\subsection{Definition of Glauber dynamics\label{sec:Glauber-dynamics}}

We now show that there is a temporal process that is able to sample
from the the distribution $\PP(\yss\cond\hh,\jj)$ \prettyref{eq:pairwise-maxent}.
This temporal dynamics is called \emph{Glauber dynamics}. It is an
example of a Markov chain on the space of binary spins $\set{0,1}^{N}$
\citep{glauber1963d}. At each time step a spin $s_{i}$ is chosen
randomly and updated with the update rule

\begin{eqnarray}
s_{i} & \leftarrow & 1\text{ with probability }F_{i}(s)=g(\sum_{j}\varLambda_{ij}s_{j}+\mu_{i})\text{ and }0\text{ else}\label{eq:Glauber}\\
g(x) & = & \frac{1}{1+\exp(-x)},\label{eq:g(x)}
\end{eqnarray}
where the coupling is assumed to be symmetric, $\varLambda_{ij}=\varLambda_{ji}$,
and self-coupling is absent, $\varLambda_{ii}=0$. The transition
operator of the Markov chain, $\kappa$, only connects states that
differ by at most one spin, so for the transition of spin $i$ we
can write, if $s^{i+}=(s_{1},\ldots,\underbrace{1}_{i-th},\ldots,s_{N})$
and $s^{i-}=(s_{1},\ldots,\underbrace{0}_{i-th},\ldots,s_{N})$, 
\begin{eqnarray}
\kappa(s^{i+}|s^{i-}) & = & F_{i}(s^{i-})\label{eq:Glauber_update}\\
\kappa(s^{i-}|s^{i+}) & = & 1-F_{i}(s^{i+}).\nonumber 
\end{eqnarray}
The pairwise maximum-entropy distribution $\PP(\yss\cond\hh,\jj)$
is stationary under the Markov dynamics above. The proof can be obtained
as the $\yji=0$ case of the proof, given below, for the inhibited
pairwise maximum-entropy model.

\subsection{Inhibited Glauber dynamics and its stationary maximum-entropy distribution\label{sec:inhibited-glauber-dynamics-explained}}

\subsubsection{Inhibited Glauber dynamics.}

In the ``inhibited'' Glauber dynamics, the network of $N$ neurons
with states $s_{i}(t)$ has an additional neuron with state $\ysi(t)$.
The dynamics is determined by the following algorithm starting at
time step $t$ with states $\yss=\yss(t),\,\ysi=\ysi(t)$:
\begin{enumerate}[label=Step \arabic*.]
\item One of the $N$ units is chosen, each unit having probability $1/N$
of being the chosen one. Suppose $i$ is the selected unit.
\item The chosen unit $i$ is updated to the state $s_{i}'\coloneqq s_{i}(t+1)$
with probability
\begin{gather*}
\begin{aligned}p(s_{i}'\cond\yss,\ysi) & =\bigl(1+\exp[(1-2s_{i}')F_{i}(\yss,\ysi)]\bigr)^{-1}\\
 & =\begin{cases}
\Bigl[1+\e^{F_{i}(\yss,\ysi)}\Bigr]^{-1}, & \text{for \ensuremath{s_{i}'=0,}}\\
\Bigl[1+\e^{-F_{i}(\yss,\ysi)}\Bigr]^{-1}, & \text{for \ensuremath{s_{i}'=1,}}
\end{cases}
\end{aligned}
\\
\text{with}\quad F_{i}(\yss,\ysi)\coloneqq\mu_{i}+\sum_{k}^{k\ne i}\varLambda_{ik}s_{k}/2+\yji\ysi.
\end{gather*}
Note the additional coupling from the neuron $\ysi$, with strength
$\yji$. This strength can have any sign, but we are interested in
the $\yji\leqslant0$ case; we therefore call $\ysi$ the ``inhibitory
neuron''.
\item The inhibitory neuron is deterministically updated to the state $\ysi'\coloneqq\ysi(t+1)$
given by
\begin{equation}
\ysi'=\HS\Bigl(\sum_{k}s_{k}/N-\theta\Bigr),
\end{equation}
corresponding to a Kronecker-delta conditional probability
\begin{equation}
p(\ysi'\cond\yss,\ysi)=p(\ysi'\cond\yss)=\dirac\bigl[\ysi'-\HS\bigl(\sum\nolimits _{k}s_{k}/N-\theta\bigr)\bigr].
\end{equation}
In other words, the inhibitory neuron becomes active if the population-averaged
activity of the other neurons is equal to or exceeds the threshold
$\theta$.
\item The time is stepped forward, $t+1\to t$, and the process repeats
from step 1.
\end{enumerate}
The original Glauber dynamics, described in the previous section,
is recovered when $\yji=0$, which corresponds to decoupling the inhibitory
neuron $\ysi$.

The total transition probability can be written as
\begin{multline}
p(\yss',\ysi'\cond\yss,\ysi)=\frac{1}{N}\,\delta\bigl[\ysi'-\HS\bigl(\sum\nolimits _{k}s_{k}/N-\theta\bigr)\bigr]\times{}\\
\sum_{i}\biggl[\bigl(1+\exp[(1-2s_{i}')F_{i}(\yss,\ysi)]\bigr)^{-1}\,\prod_{k}^{k\ne i}\dirac(s_{k}'-s_{k})\biggr];\label{eq:total_transition}
\end{multline}
the product of Kronecker deltas in the last term ensures that at most
one of the $N$ neurons changes state at each timestep.

The transition probabilities for the chosen neuron $s_{i}$ and the
inhibitory neuron $\ysi$ are independent, conditional on the state
of the network at the previous timestep:
\[
p(\yss',\ysi'\cond\yss,\ysi)=p(\yss'\cond\yss)\,p(\ysi'\cond\yss),
\]
so the transition probability for the $N$ neurons only can be written
as
\begin{gather}
p(\yss'\cond\yss)=\frac{1}{N}\,\sum_{i}\biggl[\bigl(1+\exp[(1-2s_{i}')F_{i}(\yss)]\bigr)^{-1}\,\prod_{k}^{k\ne i}\dirac(s_{k}'-s_{k})\biggr],\label{eq:transition-probability}\\
\text{with}\quad F_{i}(\yss)\coloneqq\mu_{i}+\sum_{k}^{k\ne i}\varLambda_{ik}s_{k}/2+\yji\,\HS\bigl(\sum\nolimits _{k}s_{k}/N-\theta\bigr)\bigr].\label{eq:activation-function-modified-dynamics}
\end{gather}

\subsubsection{Proof that the inhibited maximum-entropy model is the stationary
distribution of the inhibited Glauber dynamics.}

The modified maximum-entropy distribution $\PI$,  \prettyref{eq:inhibited_pairwise_maxent},
is the stationary distribution of a slightly modified version of the
above dynamics, with the update rule
\begin{equation}
\ysi'=\HS\Bigl(\sum_{k}^{k\ne i}s_{k}/N-\theta\Bigr),
\end{equation}
and the use of $N$ inhibitory neurons, one for each of the original
$N$ units. This dynamics has a slightly different transition probability,
with activation function
\begin{equation}
F_{i}(\yss)\coloneqq\mu_{i}+\sum_{k}^{k\ne i}\varLambda_{ik}s_{k}/2+\yji\,\HS\bigl(\sum\nolimits _{k}^{k\ne i}s_{k}/N-\theta\bigr)\bigr]\label{eq:modified-activation-function}
\end{equation}
instead of \prettyref{eq:activation-function-modified-dynamics}.
Note that the two dynamics are very similar for large enough $N$.
To prove the stationarity of inhibited maximum-entropy distribution
$\PI$, we show that $\PI$ satisfies the detailed-balance equality
\begin{equation}
p(\yss'\cond\yss)\,\PI(\yss)=p(\yss\cond\yss')\,\PI(\yss')\text{\quad or\quad}\frac{p(\yss'\cond\yss)}{p(\yss\cond\yss')}=\frac{\PI(\yss')}{\PI(\yss)},\quad\forall\yss,\yss',\label{eq:detailed-balance}
\end{equation}
which is a sufficient condition for stationarity \citep{kelly1979,vankampen1981_r2007,gardiner1983_r2004}.

First note that if $\yss'$ and $\yss$ differ in the state of more
than one neuron, the transition probability $p(\yss'\cond\yss)$ vanishes
and the detailed-balance above is trivially satisfied. Also the case
$\yss'=\yss$ is trivially satisfied. Only the case in which $\yss'$
and $\yss$ differ in the state of one unit, say $s_{i}$, remains
to be proven. Assume then that
\begin{equation}
s_{i}'=1,\quad s_{i}=0,\quad\forall k\ne i,\,s_{k}'=s_{k};\label{eq:last-case-detailed-balance}
\end{equation}
by symmetry, if the detailed balance is satisfied in the case above
it will also be satisfied with the values $0$ and $1$ interchanged.

Substituting the transition probability \prettyref{eq:transition-probability}
and \prettyref{eq:modified-activation-function} in the left-hand
side of the fraction form of the detailed balance \prettyref{eq:detailed-balance},
and noting that $F_{i}(\yss')=F_{i}(\yss)$, we have
\begin{equation}
\begin{split}\frac{p(\yss'\cond\yss)}{p(\yss\cond\yss')} & =\exp[-F_{i}(\yss)]^{-1}\\
 & =\exp\Bigl[\mu_{i}+\sum_{k}^{k\ne i}\varLambda_{ik}s_{k}/2+\yji\,\HS\bigl(\sum\nolimits _{k}^{k\ne i}s_{k}/N-\theta\bigr)\Bigr].
\end{split}
\label{eq:LHS-detailed-balance}
\end{equation}
Using the expression for the inhibited model $\PI$, \prettyref{eq:inhibited_pairwise_maxent},
in the right-hand side of the fraction form of the detailed balance
\prettyref{eq:detailed-balance}, we have
\begin{multline}
\frac{P(\yss')}{P(\yss)}=\exp\Bigl[\mu_{i}+\sum\limits _{k}^{k\ne i}\mu_{k}s_{k}+\tfrac{1}{2}\sum\limits _{k}^{k\ne i}\varLambda_{ik}s_{k}+{}\\
\tfrac{1}{2}\sum\limits _{k,m\ne i}^{k<m}\varLambda_{mk}s_{m}s_{k}+\yji N\,\yG\Bigl(\sum\limits _{k}^{k\ne i}\frac{s_{k}}{N}+\frac{1}{N}-\theta\Bigr)\Bigr]\\
{}=\exp\Bigl[\mu_{i}+\tfrac{1}{2}\sum\limits _{k}^{k\ne i}\varLambda_{ik}s_{k}+\yji\,\HS\bigl(\sum\limits _{k}^{k\ne i}s_{k}/N-\theta\bigr)\Bigr],\label{eq:RHS-detailed-balance}
\end{multline}
where we have used the equality $N\yG(x+1/N)-N\yG(x)=\HS(x)$, valid
if $x=\sum_{k}^{k\ne i}s_{k}/N-\theta$ and $N\theta\in\mathbf{Z}.$
Comparison of formulae \prettyref{eq:LHS-detailed-balance} and \prettyref{eq:RHS-detailed-balance}
finally proves that the detailed balance is satisfied also in the
case \prettyref{eq:last-case-detailed-balance}.

\subsection{Simulation of Glauber dynamics with NEST }

The neuron model \texttt{ginzburg\_neuron} in NEST implements the
Glauber dynamics, if the parameters of the gain function are chosen
appropriately. The gain function has the form
\begin{equation}
g_{\text{ginzburg}}(h)=c_{1}h+\frac{c_{2}}{2}(1+\tanh(c_{3}(h-\theta)).
\end{equation}
 With $\tanh(x)=\frac{e^{x}-e^{-x}}{e^{x}+e^{-x}}$, setting $x=c_{3}(h-\theta)$,
$c_{1}=0$, $c_{2}=1$, $c_{3}=\frac{1}{2}$ it takes the form
\begin{equation}
\begin{split}g_{\text{ginzburg}}(h) & =\frac{1}{2}\frac{e^{x}+e^{-x}+e^{x}-e^{-x}}{e^{x}+e^{-x}},\\
 & =\frac{1}{1+e^{-2x}}=\frac{1}{1+e^{-(h-\theta)}},
\end{split}
\end{equation}
which is identical to \prettyref{eq:g(x)}.

\section*{Acknowledgements}

We are grateful to Alexa Riehle and Thomas Brochier for providing
the experimental data and to Sonja Grün for fruitful discussions on
their interpretation. The work was carried out in the framework of
the joint International Associated Laboratory (LIA) of INT (CNRS,
AMU), Marseilles and INM-6, Jülich. Partially supported by HGF young
investigator’s group VH-NG- 1028, Helmholtz portfolio theme SMHB,
and EU Grant 604102 (Human Brain Project, HBP). All network simulations
carried out with NEST (http://www.nest-simulator.org).

PGLPM thanks the Forschungszentrum librarians for their always prompt
and efficient help in finding arcane scientific works, Miri \& Mari
for encouragement and affection, Buster for filling life with awe
and inspiration, and the developers and maintainers of \LaTeX, Emacs,
AUC\TeX, MiK\TeX, Lyx, Inkscape, bioRxiv, arXiv, HAL, PhilSci, Sci-Hub
for making a free and unfiltered science possible. 

\footnotesize\bibliographystyle{unsrtnat}
\bibliography{MaxEnt_manuscript_newARXIV}

\end{document}